%Paper: chao-dyn/9501015
%From: Giovanni Gallavotti <giovanni@ipparco.roma1.infn.it>
%Date: Sat, 28 Jan 1995 13:46:52 GMT

\newcount\mgnf\newcount\tipi\newcount\tipoformule\newcount\driver
\newcount\indice
\driver=1        %dvips
\mgnf=0          %ingrandimento
\tipi=2          %uso caratteri: 2=cmcompleti, 1=cmparziali, 0=amparziali
\tipoformule=0   %=0 da numeroparagrafo.numeroformula; se no numero
                 %assoluto
\indice=1        %=1 prepara ma non compila un nuovo indice; altro
                 %usa il vecchio

%%%%%%%%%%%%%%%%%%%%%%%%%%%%%%%%%%%%%%%%%% INCIPIT
\ifnum\mgnf=0
   \magnification=\magstep0
   \hsize=14truecm\vsize=20.truecm
   \parindent=0.3cm\baselineskip=0.45cm\fi
\ifnum\mgnf=1
   \magnification=\magstep1\hoffset=0.truecm
   \hsize=14truecm\vsize=20truecm
   \baselineskip=18truept plus0.1pt minus0.1pt \parindent=0.9truecm
   \lineskip=0.5truecm\lineskiplimit=0.1pt      \parskip=0.1pt plus1pt\fi
%%%%%%%%%%%%%%%%%%%%%%%%%%%%%%%%%%%%%%%%%%
%%%%%%%%%%%%%%%%%%%%%%%%%%% GRECO
%%%%%%%%%%%%%%%%%%%%%%%%%%%%%%%%%%%%%%%%%%%%%%%%
\let\a=\alpha \let\b=\beta  \let\g=\gamma     \let\d=\delta  \let\e=\varepsilon
\let\z=\zeta  \let\h=\eta   \let\th=\vartheta    \let\l=\lambda
\let\m=\mu    \let\n=\nu    \let\x=\xi        \let\p=\pi      \let\r=\rho
\let\s=\sigma \let\t=\tau   \let\f=\varphi     
               \let\o=\omega

 \let\D=\Delta     \let\L=\Lambda  
\let\P=\Pi     \let\F=\Phi

%%%%%%%%%%%%%%%%%%%%%%%%%%%%%%%%%%%%%%%%%%%%%%%%
%%%%%%%%%%%%%%%%%%%%%%%%%%%%%%%%%%%%%%%%%%%%%%%%%%%%%%%%%%%%%%
%%%%%%%%%%%%%%%%% EQUAZIONI CON NOMI SIMBOLICI
%%%
%%% per assegnare un nome simbolico ad una equazione basta
%%% scrivere \Eq(...) o, in \eqalignno, \eq(...) o,
%%% nelle appendici, \Eqa(...) o \eqa(...):
%%% dentro le parentesi e al posto dei ...
%%% si puo' scrivere qualsiasi commento;
%%% per assegnare un nome simbolico ad una figura, basta scrivere
%%% \geq(...); per avere i nomi
%%% simbolici segnati a sinistra delle formule e delle figure si deve
%%% dichiarare il documento come bozza, iniziando il testo con
%%% \BOZZA. Sinonimi \Eq,\EQ,\EQS; \eq,\eqs; \Eqa,\Eqas;\eqa,\eqas.
%%% All' inizio di ogni paragrafo si devono definire il
%%% numero del paragrafo e della prima formula dichiarando
%%% \numsec=... \numfor=...  (brevetto Eckmannn); all'inizio del lavoro
%%% bisogna porre \numfig=1 (il numero delle figure non contiene la sezione.
%%% Si possono citare formule o figure seguenti; le corrispondenze fra nomi
%%% simbolici e numeri effettivi sono memorizzate nel file \jobname.aux, che
%%% viene letto all'inizio, se gia' presente. E' possibile citare anche
%%% formule o figure che appaiono in altri file, purche' sia presente il
%%% corrispondente file .aux; basta includere all'inizio l'istruzione
%%%           \include{nomefile}
%%%%%%%%% NOTA BENE
%%% Per il buon funzionamento dei riferimenti occorre togliere il
%%% commento della riga \openout15=\jobname.aux ORA COMMENTATA,
%%% e compilare DUE volte.
%%%%%%%%%%%%%%%%%%%%%%%%%%%%%%%%%%%%%%%%%%%%%%%%%%%%%%%%%%%%%%%

\global\newcount\numsec\global\newcount\numfor
\global\newcount\numapp\global\newcount\numcap
\global\newcount\numfig\global\newcount\numpag
\global\newcount\numnf

\def\SIA #1,#2,#3 {\senondefinito{#1#2}%
\expandafter\xdef\csname #1#2\endcsname{#3}\else
\write16{???? ma #1,#2 e' gia' stato definito !!!!} \fi}

\def \FU(#1)#2{\SIA fu,#1,#2 }

\def\etichetta(#1){(\veroparagrafo.\veraformula)%
\SIA e,#1,(\veroparagrafo.\veraformula) %
\global\advance\numfor by 1%
%\write15{\string\FU (#1){\equ(#1)}}%
\write16{ EQ #1 ==> \equ(#1)  }}
\def\etichettaa(#1){(A\veraappendice.\veraformula)
 \SIA e,#1,(A\veraappendice.\veraformula)
 \global\advance\numfor by 1
% \write15{\string\FU (#1){\equ(#1)}}
 \write16{ EQ #1 ==> \equ(#1) }}
\def\getichetta(#1){Fig. \verafigura
 \SIA g,#1,{\verafigura}
 \global\advance\numfig by 1
% \write15{\string\FU (#1){\graf(#1)}}
 \write16{ Fig. #1 ==> \graf(#1) }}
\def\retichetta(#1){\numpag=\pgn\SIA r,#1,{\verapagina}
% \write15{\string\FU (#1){\rif(#1)}}
 \write16{\rif(#1) ha simbolo  #1  }}
\def\etichettan(#1){(n\verocapitolo.\veranformula)
 \SIA e,#1,(n\verocapitolo.\veranformula)
 \global\advance\numnf by 1
\write16{\equ(#1) <= #1  }}

\newdimen\gwidth
\gdef\profonditastruttura{\dp\strutbox}
\def\senondefinito#1{\expandafter\ifx\csname#1\endcsname\relax}
\def\BOZZA{
\def\alato(##1){
 {\vtop to \profonditastruttura{\baselineskip
 \profonditastruttura\vss
 \rlap{\kern-\hsize\kern-1.2truecm{$\scriptstyle##1$}}}}}
\def\galato(##1){ \gwidth=\hsize \divide\gwidth by 2
 {\vtop to \profonditastruttura{\baselineskip
 \profonditastruttura\vss
 \rlap{\kern-\gwidth\kern-1.2truecm{$\scriptstyle##1$}}}}}
\def\verapagina{
{\romannumeral\number\numcap}.\number\numsec.\number\numpag}}

\def\alato(#1){}
\def\galato(#1){}
\def\veroparagrafo{\number\numsec}\def\veraformula{\number\numfor}
\def\veraappendice{\number\numapp}
\def\verapagina{\number\pageno}\def\veranformula{\number\numnf}
\def\verafigura{{\romannumeral\number\numcap}.\number\numfig}
\def\verocapitolo{\number\numcap}\def\veranformula{\number\numnf}
\def\Eqn(#1){\eqno{\etichettan(#1)\alato(#1)}}
\def\eqn(#1){\etichettan(#1)\alato(#1)}

\def\Eq(#1){\eqno{\etichetta(#1)\alato(#1)}}
\def\eq(#1){\etichetta(#1)\alato(#1)}
\def\Eqa(#1){\eqno{\etichettaa(#1)\alato(#1)}}
\def\eqa(#1){\etichettaa(#1)\alato(#1)}
\def\dgraf(#1){\getichetta(#1)\galato(#1)}
\def\drif(#1){\retichetta(#1)}

\def\eqv(#1){\senondefinito{fu#1}$\clubsuit$#1\else\csname fu#1\endcsname\fi}
\def\equ(#1){\senondefinito{e#1}\eqv(#1)\else\csname e#1\endcsname\fi}
\def\graf(#1){\senondefinito{g#1}\eqv(#1)\else\csname g#1\endcsname\fi}
\def\rif(#1){\senondefinito{r#1}\eqv(#1)\else\csname r#1\endcsname\fi}
%%%%%%%%%%%%%%%%%%%%%%%%%%%%%%%%%%%%%%%%%%%%%%%%%%%%%%%%%%%%%%
%%%%%%%%%%%%%%%%%% Numerazione verso il futuro ed eventuali paragrafi
%%%%%%%%%%%%%%%%%% precedenti non inseriti nella scheda da compilare
%%%%%%%%%%%%%%%%%% e elenco referenze bibliografiche creato in
%%%%%%%%%%%%%%%%%% \jobname.bib

%\openin14=\jobname.aux \ifeof14 \relax \else
%\input \jobname.aux \closein14 \fi
%14 e' libero !!

%%%%%%%%%%%%%%%%%%%%%%%%%%%%

%%%%%%%%%%%%%%%%%%%%%%%%%%%%%%%%%%%%%%%%%%%%%%%%%%%%%%%%%%%%%%

%\newcount\tipoformule
%\tipoformule=1   %=0 da numeroparagrafo.numeroformula; se no numero
%                 %assegnato
\ifnum\tipoformule=1\let\Eq=\eqno\def\eq{}\let\Eqa=\eqno\def\eqa{}
\def\equ{}\fi

%%%%%%%%%%%%%%%%%%%%%%%%%%%%%%%%%%%%%%%%%%
%%%%%%%%%%%%%%%%%%%%%%%%%%%%%%%%%%%%%%%%%%%%%%%%%%%%
%%%%%%%%%%%%%%%%%%%%%  Numerazione pagine

{\count255=\time\divide\count255 by 60 \xdef\hourmin{\number\count255}
	\multiply\count255 by-60\advance\count255 by\time
   \xdef\hourmin{\hourmin:\ifnum\count255<10 0\fi\the\count255}}

\def\oramin{\hourmin }

\def\data{\number\day/\ifcase\month\or gennaio \or febbraio \or marzo \or
aprile \or maggio \or giugno \or luglio \or agosto \or settembre
\or ottobre \or novembre \or dicembre \fi/\number\year;\ \oramin}

\setbox200\hbox{$\scriptscriptstyle \data $}

\newcount\pgn \pgn=1
\def\foglio{\number\numsec:\number\pgn
\global\advance\pgn by 1}
\def\foglioa{A\number\numsec:\number\pgn
\global\advance\pgn by 1}

%\footline={\rlap{\hbox{\copy200}\ $\st[\number\pageno]$}\hss\tenrm
%\foglio\hss}
%\footline={\rlap{\hbox{\copy200}\ $\st[\number\pageno]$}\hss\tenrm
%\foglioa\hss}
%

%\footline={\rlap{\hbox{\copy200}\ $\st[\number\pageno]$}\hss\tenrm\foglio\hss}
%\footline={\rlap{\hbox{\copy200}}\hss\tenrm\folio\hss}
\footline={\rlap{\hbox{\copy200}}}

%%%%%%%%%%%%POSTSCRIPT
%
% Inizializza le macro postscript e il tipo di driver di stampa.
% Attualmente le istruzioni postscript vengono utilizzate solo se il driver
% e' DVILASER ( \driver=0 ), DVIPS ( \driver=1) o PSPRINT ( \driver=2);
% qualunque altro valore di \driver produce un output in cui le figure
% contengono solo i caratteri inseriti con istruzioni TEX (vedi avanti).
%
\newdimen\xshift \newdimen\xwidth
%
% inserisce una scatola contenente #3 in modo che l'angolo superiore sinistro
% occupi la posizione (#1,#2)
%
\def\ins#1#2#3{\vbox to0pt{\kern-#2 \hbox{\kern#1 #3}\vss}\nointerlineskip}
%
% Crea una scatola di dimensioni #1x#2 contenente il disegno descritto in
% #4.ps; in questo disegno si possono introdurre delle stringhe usando \ins
% e mettendo le istruzioni relative al #3 (che puo' anche mancare);
% al disotto del disegno, al centro, e' inserito il numero della figura
% calcolato tramite \geq(#4).
% Il file #4.ps contiene le istruzioni postscript, che devono essere scritte
% presupponendo che l'origine sia nell'angolo inferiore sinistro della
% scatola, mentre per il resto l'ambiente grafico e' quello standard.
% Se \driver=2, e' necessario dilatare la figura in accordo al valore di
% \magnification, correggendo i parametri P1 e P2 nell'istruzione
%         \special{#4.pst P1 P2 scale}
%
\def\insertplot#1#2#3#4{
    \par \xwidth=#1 \xshift=\hsize \advance\xshift
     by-\xwidth \divide\xshift by 2 \vbox{
  \line{} \hbox{ \hskip\xshift  \vbox to #2{\vfil
 \ifnum\driver=0 #3
                 \special{ps: plotfile #4.ps}
 \fi
 \ifnum\driver=1  #3    \includegraphics{#4.ps}       \fi
 \ifnum\driver=2  #3   \ifnum\mgnf=0
                       \special{#4.ps 1. 1. scale}\fi
                       \ifnum\mgnf=1
                       \special{#4.ps 1.2 1.2 scale}\fi\fi
 \ifnum\driver=5  #3   \fi}
\hfil}}}

\newdimen\xshift \newdimen\xwidth \newdimen\yshift
\def\eqfig#1#2#3#4#5{
\par\xwidth=#1 \xshift=\hsize \advance\xshift
by-\xwidth \divide\xshift by 2
\yshift=#2 \divide\yshift by 2
\line{\hglue\xshift \vbox to #2{\vfil
\ifnum\driver=0 #3
%
\special{ps: plotfile #4.ps}
\fi
\ifnum\driver=1 #3 \includegraphics{#4.ps}\fi
\ifnum\driver=2 #3 \special{
\ifnum\mgnf=0 #4.ps 1. 1. scale \fi
\ifnum\mgnf=1 #4.ps 1.2 1.2 scale\fi}
\fi}\hfill\raise\yshift\hbox{#5}}}

\def\figini#1{
\def\8{\write13}
\catcode`\%=12\catcode`\{=12\catcode`\}=12
\catcode`\<=1\catcode`\>=2
\openout13=#1.ps}

\def\figfin{
\closeout13
\catcode`\%=14\catcode`\{=1
\catcode`\}=2\catcode`\<=12\catcode`\>=12
}

%%%%%%%%%%%%%%%%%%%%%%%%%%%%%%%%%%%%%%%%%%%%% CARATTERI %%%%%%%%%%%%%%
\newskip\ttglue
%% am
\def\TIPIO{
\font\setterm=amr7 %\font\settei=ammi7
\font\settesy=amsy7 \font\settebf=ambx7 %\font\setteit=amit7
%%%%% cambiamenti di formato %%%
\def \settepunti{\def\rm{\fam0\setterm}% passaggio a tipi da 7-punti
\textfont0=\setterm   %\textfont1=\settei
\textfont2=\settesy   %\textfont3=\setteit
%\textfont\itfam=\setteit  \def\it{\fam\itfam\setteit}
\textfont\bffam=\settebf  \def\bf{\fam\bffam\settebf}
\normalbaselineskip=9pt\normalbaselines\rm
}\let\nota=\settepunti}
\def\annota#1#2{\footnote{${}^#1$}{\nota#2\vfill}}
%%%%%%%

%%cm completo
\def\TIPITOT{
\font\twelverm=cmr12
\font\twelvei=cmmi12
\font\twelvesy=cmsy10 scaled\magstep1
\font\twelveex=cmex10 scaled\magstep1
\font\dodici=cmbx10 scaled\magstep1
\font\twelveit=cmti12
\font\twelvett=cmtt12
\font\twelvebf=cmbx12
\font\twelvesl=cmsl12
\font\ninerm=cmr9
\font\ninesy=cmsy9
\font\eightrm=cmr8
\font\eighti=cmmi8
\font\eightsy=cmsy8
\font\eightbf=cmbx8
\font\eighttt=cmtt8
\font\eightsl=cmsl8
\font\eightit=cmti8
\font\sixrm=cmr6
\font\sixbf=cmbx6
\font\sixi=cmmi6
\font\sixsy=cmsy6
%%%%%%%%%%%%%%%%%%%%%%%%%%%%%%%%%%%%%%%
\font\twelvetruecmr=cmr10 scaled\magstep1
\font\twelvetruecmsy=cmsy10 scaled\magstep1
\font\tentruecmr=cmr10
\font\tentruecmsy=cmsy10
\font\eighttruecmr=cmr8
\font\eighttruecmsy=cmsy8
\font\seventruecmr=cmr7
\font\seventruecmsy=cmsy7
\font\sixtruecmr=cmr6
\font\sixtruecmsy=cmsy6
\font\fivetruecmr=cmr5
\font\fivetruecmsy=cmsy5
%%%% definizioni per 10pt %%%%%%%%
\textfont\truecmr=\tentruecmr
\scriptfont\truecmr=\seventruecmr
\scriptscriptfont\truecmr=\fivetruecmr
\textfont\truecmsy=\tentruecmsy
\scriptfont\truecmsy=\seventruecmsy
\scriptscriptfont\truecmr=\fivetruecmr
\scriptscriptfont\truecmsy=\fivetruecmsy
%%%%% cambio grandezza %%%%%%
\def \eightpoint{\def\rm{\fam0\eightrm}% switch to 8-point type
\textfont0=\eightrm \scriptfont0=\sixrm \scriptscriptfont0=\fiverm
\textfont1=\eighti \scriptfont1=\sixi   \scriptscriptfont1=\fivei
\textfont2=\eightsy \scriptfont2=\sixsy   \scriptscriptfont2=\fivesy
\textfont3=\tenex \scriptfont3=\tenex   \scriptscriptfont3=\tenex
\textfont\itfam=\eightit  \def\it{\fam\itfam\eightit}%
\textfont\slfam=\eightsl  \def\sl{\fam\slfam\eightsl}%
\textfont\ttfam=\eighttt  \def\tt{\fam\ttfam\eighttt}%
\textfont\bffam=\eightbf  \scriptfont\bffam=\sixbf
\scriptscriptfont\bffam=\fivebf  \def\bf{\fam\bffam\eightbf}%
\tt \ttglue=.5em plus.25em minus.15em
\setbox\strutbox=\hbox{\vrule height7pt depth2pt width0pt}%
\normalbaselineskip=9pt
\let\sc=\sixrm  \normalbaselines\rm
\textfont\truecmr=\eighttruecmr
\scriptfont\truecmr=\sixtruecmr
\scriptscriptfont\truecmr=\fivetruecmr
\textfont\truecmsy=\eighttruecmsy
\scriptfont\truecmsy=\sixtruecmsy
}\let\nota=\eightpoint}

\newfam\msbfam   %per uso in \TIPITOT
\newfam\truecmr  %per uso in \TIPITOT
\newfam\truecmsy %per uso in \TIPITOT
%%%%%%%%%%%%%%%%%%%%%%%%%%%%%%%
%%Scelta dei caratteri
%\newcount\tipi \tipi=0   %e' definito all'inizio
\newskip\ttglue
\ifnum\tipi=0\TIPIO \else\ifnum\tipi=1 \TIPI\else \TIPITOT\fi\fi

%%%%%%%%%%%%%%%%%%%%%%%%%%%%%% DEFINIZIONI LOCALI
%%%%%%%%%%%%%%%%%%%%%%%%%%%%%%%%%%%%%%%%%%%%%%%%%%%%

\let\0=\noindent

\def\media#1{{\langle#1\rangle}}
\def\ie{\hbox{\it i.e.\ }}\def\eg{\hbox{\it e.g.\ }}
\let\dpr=\partial\def\\{\hfill\break}

\def\*{\vglue0.3truecm}\let\0=\noindent
\let\==\equiv
\let\txt=\textstyle

\def\mbe{{\\*\hfill\hbox{\it
mbe\kern0.5truecm}}\vskip3.truept}
\def\1{{-1}}
\let\io=\infty \def\V#1{\,\vec#1}   \def\Dpr{\V\dpr\,}
    \let\ig=\int

\def\tende#1{\,\vtop{\ialign{##\crcr\rightarrowfill\crcr
              \noalign{\kern-1pt\nointerlineskip}
              \hskip3.pt${\scriptstyle #1}$\hskip3.pt\crcr}}\,}
\def\otto{\,{\kern-1.truept\leftarrow\kern-5.truept\to\kern-1.truept}\,}
\def\fra#1#2{{#1\over#2}}

\global\newcount\numpunt
\def\XWPR{{\it a priori}}
\def\ap#1{\def\9{#1}{\if\9.\global\numpunt=1\else\if\9,\global\numpunt=2\else
\if\9;\global\numpunt=3\else\if\9:\global\numpunt=4\else
\if\9)\global\numpunt=5\else\if\9!\global\numpunt=6\else
\if\9?\global\numpunt=7\else\global\numpunt=8\fi\fi\fi\fi\fi\fi
\fi}\ifcase\numpunt\or{\XWPR.}\or{\XWPR,}\or
{\XWPR;}\or{\XWPR:}\or{\XWPR)}\or
{\XWPR!}\or{\XWPR?}\or{\XWPR\ \9}\else\fi}
%%%%%%%%%%%%%%%%%%%%%%%%

\def\fiat{{}}
%%%%%%%%%%%%%%%%%%%%%%%%%%%%%%%%%%%%%%%%%%
%%%DEFINIZIONI LOCALI
    \def\kk{{\V k}}

\def\V#1{{\underline#1}}
\def\2{{1\over2}}
\def\EE{{\cal E}}
\def\CC{{\cal C}}\def\FF{{\cal F}}

\def\T#1{{#1_{\kern-3pt\lower7pt\hbox{$\widetilde{}$}}\kern3pt}}
\def\VV#1{{\underline #1}_{\kern-3pt
\lower7pt\hbox{$\widetilde{}$}}\kern3pt\,}
\def\W#1{#1_{\kern-3pt\lower7.5pt\hbox{$\widetilde{}$}}\kern2pt\,}

\def\NN{{\cal N}}
\def\LL{{\cal L}}
\def\lis{\overline}

\def\mod{{\rm mod}\,}

%%%%%%%%%%%%%%%%%%%%%%%%%%%%%%%%%%%%%%

\def\indica{\leaders \hbox to 0.5cm{\hss.\hss}\hfill}

%\openout15=\jobname.aux        %\write15
%%%%%%%%%%%%%%%%%%%%%%%%%%%%%%%%% Introduzione

\def\guida{\leaders\hbox to 1em{\hss.\hss}\hfill}

\let\h=\eta\let\x=\xi\def\ie{{\it i.e. }}   
\newtoks\footline \footline={\hss\tenrm\folio\hss}
%\headline{\hss\tenrm \it DRAFT 4, not for circulation}
\footline={\rlap{\hbox{\copy200}}\tenrm\hss \number\pageno\hss}
\let\ciao=\bye
%\BOZZA

\def\annota#1#2{\footnote{${}^#1$}{\nota#2\vfill}}

\def\equ{{}}\let\Eq=\eqno\let\eq=\eqno
%\headline{\nota\hss DRAFT 19: NOT FOR CIRCULATION}
%\BOZZA
\fiat

\vglue2cm \centerline{\twelvebf Dynamical ensembles in stationary
states.\annota{!}{Archived in {\it mp$\_$arc@ math. utexas. edu}, \#95-32}}
\vskip1.cm

\centerline{ G. Gallavotti\footnote{${}^\#$}{\nota
Fisica, Universit\`a di Roma La Sapienza, p.le Moro 2, 00185, Roma,
Italia.}, E. G. D. Cohen\footnote{${}^*$}{\nota The Rockefeller
University, New York, N.Y. 10021, USA.}}
\vskip2cm

\0{\it Abstract: We propose as a generalization of an idea of Ruelle to
describe turbulent fluid flow a chaotic hypothesis for reversible
dissipative many particle systems in nonequilibrium stationary states in
general. This implies an extension of the zeroth law of thermodynamics
to non equilibrium states and it leads to the identification of a unique
distribution $\m$ describing the asymptotic properties of the time
evolution of the system for initial data randomly chosen with respect to
a uniform distribution on phase space. For conservative systems in
thermal equilibrium the chaotic hypothesis implies the ergodic
hypothesis. We outline a procedure to obtain the distribution $\m$: it
leads to a new unifying point of view for the phase space behavior of
dissipative and conservative systems. The chaotic hypothesis is
confirmed in a non trivial, parameter--free, way by a recent computer
experiment on the entropy production fluctuations in a shearing fluid
far from equilibrium. Similar applications to other models are proposed,
in particular to a model for the Kolmogorov--Obuchov theory for
turbulent flow.}

\vskip1cm
\*

\0{\it\S1 Introduction.}
\numsec=1\numfor=1
\*

In a previous paper [CG] we proposed the use of Ruelle's idea (discussed
in \S2) to obtain the probability distribution for the statistics of
turbulent flows in hydrodynamics, as a basis for the study of many
particle statistical mechanical systems in nonequilibrium stationary
states in general. We did so, by providing a concrete procedure of how
to obtain the necessary probability distribution, now called the
Sinai--Ruelle--Bowen (SRB) distribution, to compute the statistical
properties of the above mentioned systems. The applicability of such a
distribution has, so far, only been proved with mathematical rigor for
very idealized systems, such as Anosov or Axiom A systems, and it would
be impossible at present to give extensions of the proofs for the many
particle systems of interest here. Therefore we proposed to use Ruelle's
idea as a heuristic principle to obtain the statistical properties of
such systems, at least when they are very large, \ie in the
thermodynamic limit. This implied that we made a ``chaotic hypothesis''
that the many particle systems in statistical mechanics are essentially
chaotic in the sense of Anosov, \ie they behave {\it as if they were}
Anosov systems as far as their properties of physical interest are
concerned. In other words we use the SRB distribution obtained from the
strong assumption of chaoticity in the Anosov sense in a {\it heuristic}
way to compute statistical mechanical properties of our system and
assume that the corrections due to the possible non--validity of the
strong chaoticity assumption become negligible for large systems.

The position that we take here is very similar to that usually taken
with respect to the so--called ergodic hypothesis, which has been proven
only for very special few particle systems. Yet, when used as a
principle, it has led to all known results of statistical mechanics,
beginning with its connection with thermodynamics. It would seem
therefore inappropriate, in fact very unfortunate, if the application of
the ergodic hypothesis, would have had to wait till it had been proved
valid for the many particle statistical mechanical systems in thermal
equilibrium of physical interest. Very recently a version of
what we shall call the {\it chaotic hypothesis}, see \S2, has been
rigorously proved for a single particle system held in a non equilibrium
stationary state and a number of detailed consequences have been
derived, which agree with experiment [CELS].

Here we will give a number of possible many particle systems to which
the chaotic hypothesis and the ensuing SRB distribution could
immediately be applied. So far, only one of those systems: a shearing
thermostatted fluid far from equilibrium (see \S3, model 2) has been
investigated experimentally, {\it viz.}  the statistics of the
fluctuations of the pressure tensor -- or equivalently of the entropy
production rate -- in this system have been determined numerically and
found to be in very good agreement with what one obtains by applying the
chaotic hypothesis. Although corresponding experiments have not been
done as yet for the other systems we mention, they should provide
further checks on the validity of Ruelle's ideas and the chaotic
hypothesis as proposed here.

We want to emphasize that the application of the chaotic hypothesis is
not restricted to stationary states near equilibrium, \ie to the linear
regime of small deviations from thermal equilibrium, as the above
mentioned example of a shearing flow shows. The precise limitations of
its applicability are unknown, however.

The way we will present the construction of the SRB distribution from
the chaotic hypothesis can also be applied to the theory of
equilibrium states. It leads then to a new picture of the behavior in phase
space of both equilibrium and stationary nonequilibrium systems, which
reveals a much closer analogy in their phase space behavior than
considered up till now. Thus a unification of the conservative behavior
in equilibrium states and of the dissipative behavior in non
equilibrium stationary states emerges.

In \S2 we describe some general properties that can help visualizing the
general phenomenology of the non equilibrium systems that we consider:
the discussion leads then to a formal definition of Ruelle's idea and to
the precise formulation of the chaotic hypothesis.  In \S3 we give a
variety of examples of nonequilibrium systems to which the chaotic
hypothesis can be straightforwardly applied. In \S4 we discuss from a
somewhat unusual viewpoint the heuristic ideas behind the hypothesis;
this leads, in \S5 and \S6, to an outline and reinterpretation of the
classical, [S2,Bo,R1], construction of the appropriate SRB distribution
for this system, using Markov partitions. In \S7 we briefly summarize
the only concrete application so far available, {\it viz.}  that of a
shearing fluid, and we discuss our main result, the fluctuation theorem
of \S7 (which gives a theoretical interpretation of the experiment).  In
\S8 we give a discussion and outlook.  \*

\0{\it\S2 The SRB picture.}  \numsec=2\numfor=1 \*

For a convenient discussion of the SRB picture of nonequilibrium
stationary states it is important to discuss the time evolution in
discrete time, rather than in continuous time. This will be obtained by
observing the motion when some timing event happens (this is usually
done by describing the motion through a Poincar\'e section).  Therefore
we fix a {\it timing event} and envisage performing our observations at
every time the event happens.  This will have the effect of reducing by
one unit the initial phase space dimension.

\0The choice of the timing event is essentially arbitrary: for many
particle systems a reasonable choice could be the event in which the
pair of closest particles (among the $N$ we have) is at a distance $ r$,
coming from larger distances.  We call such event a ``collision'' and we
use it as our timing event.  To avoid trivialities $ r$ has to be chosen
small compared to the average interparticle distance, but not too small
(\ie larger than the ``core'' of the interaction).  In the case of a
continuous fluid flow a natural timing event, actually used in many
numerical experiments starting with [Lo], is the event in which a given
coordinate of the velocity field passes through a prefixed value, or
assumes a locally maximum value. To uniformize the notations we shall
also call such an event a ``collision''.

\0The dynamical systems we consider will be defined now by the phase
space $\CC$ of the ``collisions'', with dimension $2D$ and the time
evolution, which will be a map $S: \CC\to \CC$ defined by $Sx=S_{\t(x)}
x$, if $S_t$ is the continuous time evolution operator solving the
equations of motion in the full phase space $\FF$, which in our cases
will coincide with a constant ``energy'' surface or it will be a
manifold in it. Here $\t(x)$ is the time interval between the collision
$x\in \CC$ and the next one.

We shall make a statistical study (like is done in equilibrium): this
means that we shall be interested in the properties of the time
evolutions of the motions that can be seen by extracting the initial
data at random with the Liouville distribution on $\FF$. Since our
analysis will be performed on $\CC$ rather than on $\FF$ we shall need
the corresponding probability distribution on $\CC$.  The Liouville
distribution $\m_\LL= const\, \d(H(\V p,\V q)-E)d\V p d\V q$ (or
$\m_\LL=const\,\d(\sum|\V\g_\kk|^2-E)$ in the case of the fluid motion
models that we consider in \S3, where the variables $\V\g_\kk$ are the
Fourier components of the velocity field) on the full phase space $\FF$
(energy surface) naturally generates a probability distribution $\m_0$
on $\CC$: if $E$ is a set on $\CC$ we simply set $\m_0(E)$ equal to the
Liouville measure of the tube of trajectory segments in $\FF$ starting
at $E$ and ending at the next collision, when evolved with the
motion corresponding to no external forcing fields.  We shall still call
$\m_0$ the ``Liouville distribution'' (on $\CC$).

The first point of our analysis is a generalization of the {\it zeroth
law of thermodynamics} to nonequilibrium stationary states. As expressed
by Uhlenbeck and Ford, [UF], the zeroth law of thermodynamics states
that a closed conservative mechanical system consisting of a very large
number of particles will, when initially not in equilibrium, approach
equilibrium, where all {\it macroscopic} variables have reached
stationary values. By (asymptotic) equilibrium one means here that the
time averages have reached the value that can be computed by a
probability distribution on the energy surface. This law can be {\it
extended} to nonequilibrium systems as: \*

\0{\it Extended zero-th law: A dynamical system $(\CC,S)$ describing a many
particle system (or a continuum such as a fluid) describes motions that
admit a statistics $\m$ in the sense that, given any (piecewise smooth),
macroscopic observable $F$ defined on the points $x$ of the phase space
$\CC$, the time average of $F$ exists for all $\m_0$--randomly--chosen
initial data $x$ and is given by:
$$\lim_{T\to\io}\fra1T\sum_{k=0}^{T-1}
F(S^jx)=\ig_\CC\m(dx')F(x')\Eq(2.1)$$
where $\m$ is a $S$--invariant probability distribution on $\CC$.}  \*

In this form we suppose that it holds for all our models. The notation
$\m(dx)$ rather than $r(x)dx$ expresses the possible fractal nature of
the support of the distribution $\m$, and implies that the the
probability to find the dynamical system in the infinitesimal volume
$dx$ around $x$ may not be proportional to $dx$, so that it cannot be
written as $r(x)dx$ with $r(x)$ a probability density and $dx$ the
volume measure on phase space.

The main point of this paper is to use an idea of Ruelle's as a guiding
principle to describe nonequilibrium stationary states in general.  That
is, we propose that for such systems there exists a distribution
(usually called the SRB distribution) describing the asymptotic
statistics of motions with initial data randomly chosen with respect to
a uniform distribution on phase space (the Liouville distribution).  For
this to be realistically implemented we assume that {\it macroscopic
systems, consisting of very many particles}, behave as transitive Anosov
systems, \ie are ``chaotic'' in the sense that each point $x$ in phase
space admits an unstable and a stable manifold $W^u_x,W^s_x$ which
depend continuously on $x$, are dense in the phase space $\CC$, and on
which the expansion and contraction rates are everywhere separated by a
finite gap from $0$ (hence no zero Lyapunov exponents
occur).\footnote{${}^1$}{\nota For convenience we formally recall that
an Anosov system $(\CC,S)$ is a {\it smooth} dynamical system such that
{\it every} point $x\in\CC$ possesses stable and unstable manifolds
$W^s_x,W^u_x$ which depend continuously on $x$ and on which $S^n,S^{-n}$
respectively contract infinitesimal vectors by a factor bounded by $C
e^{-\l n}$, for $n\ge0$, and likewise for $n\le0$ they expand by a
factor bounded by $C^{-1} e^{-\l n}$. The constant $\l$ is therefore
such that all Lyapunov exponents verify $|\l_j|\ge\l$ and, hence, $\l$
can be called a bound on the {\it Lyapunov spectrum gap}. Note that the
continuity of the $W^u_x,W^s_x$ in $x$ implies the {\it transversality}
of the two manifolds, which therefore form everywhere an angle $\th(x)$
bounded away from $0$ and $\p$. An Anosov system is transitive if
$W^u_x,W^s_x$ are dense in $\CC$ for all $x$.}

We propose therefore the following chaotic hypothesis, which in [CG] we
called {\it Ruelle's principle}, as a generalization of Ruelle's idea:

\* {\it Chaotic hypothesis: A reversible many particle system in a
stationary state can be regarded as a transitive Anosov system for the
purpose of computing the macroscopic properties of the system.}  \*

We intend to show that this hypothesis, although general, leads to
concrete verifiable consequences and may be, in this respect, similar to
the ergodic hypothesis for equilibrium states but, unlike the ergodic
hypothesis, admits an extension to nonequilibrium stationary states.

One could weaken our form of the chaotic hypothesis by replacing
``Anosov system'' with ``Axiom A system'', and refer to the general
theory of such systems developed in [Bo],[R2] (rather than relying on the
work of Sinai on Anosov systems, [S2]); one could even attempt to
weaken it further by trying to make use of the general theory of Pesin
of non smooth hyperbolic system, [P]. However, we shall not dwell on
such somewhat obvious extensions of our ideas, as they do not seem
relevant at present.

Examples of model systems in nonequilibrium stationary states to which
the chaotic hypothesis is applicable will be given in \S3. Numerical
evidence leads us to believe that they seem to share a number of
properties which we believe to hold also for more general physical systems
and which we now summarize. Not all of them are necessary for the
applications we shall discuss: however, they are very helpful for
building an intuitive, model independent, picture of the phenomena that
we attempt to study.  When discussing the models from a technical
viewpoint we shall mention which properties have been
experimentally checked and which have not (yet) been
checked: our applications will only require properties (A,B,C) below.
\*

\0{\it (A) Dissipation:} the phase space volume undergoes a contraction
at a rate, on the average, equal to $D\media{\s(x)}_+$ where $2D$ is the
phase space $\CC$ dimension and $\s(x)$ is a model dependent ``rate''
per degree of freedom. The average here is a time average from time zero
to plus infinity and the rate is a generalization of the usual {\it
entropy production rate}, (see \S3 for motivation of this remark).

We say that a system is {\it dissipative} if the contraction rate per
degree of freedom, $\media\s_+$, is positive. We shall assume that the
models that we consider here in nonequilibrium situations are all
dissipative. The instantaneous contraction rate $\s(x)$ is, however, a
fluctuating quantity and we note that when we consider in this paper
entropy production rates and their fluctuations we identify them,
mathematically, with phase space contraction rates and their
fluctuations, respectively.\*

\0{\it (B) Reversibility:} there is an isometry, \ie a metric preserving
map, $i$ in phase space, which is a map $i: x\to ix$ such that if $t\to
x(t)$ is a solution, then $i(x(-t))$ is also a solution, and furthermore
$i^2$ is the identity.  \*

\0{\it (C) Chaoticity:} the above chaotic hypothesis holds and we can
treat the system $(\CC,S)$ as a transitive Anosov systems.

\* We realize that (C) {\it cannot} hold strictly in finite systems,
even in the case of smooth interaction potentials (in the presence of
hard cores the Anosov property, which requires smoothness of the
dynamics as a prerequisite, is in fact obviously false). What we mean
here is that we assume that the system behaves {\it as if it was a
transitive Anosov system} and that the errors made become negligible
(even when there are hard core collisions) at least in the large system
limit.

In support of (A,B,C) the following two properties (D) and (E) also are
relevant and appear to hold at least for some of the models that we shall
treat: \*

\0{\it (D) Pairing of Lyapunov exponents:} half of the $2D$ Lyapunov
exponents are $\ge0$ and half are $<0$. If they are ordered so that
$0\le \l^+_1\le\l^+_2\le\ldots\le \l^+_D$ and $0>\l^-_1\ge
\l^-_2>\ldots>\l^-_D$ so that $\l_{\max}=\l^+_D$ the following {\it
pairing rule} holds:

$$\l^+_j+\l^-_j=\fra1D \sum_{k=1}^{2D}\l_k\=-\media{\s}_+,\qquad
j=1,\ldots,D\Eq(2.2)$$

\0which has been proved for some special cases where the system is non
reversible (\ie $\s(x)$ is constant) and the pairs do not necessarily
consist of exponents with opposite sign; it has been found  numerically,
in the case of model 2 and related models in the form \equ(2.2) in
[ECM1], [SEM] (where it was formulated in the above form).  \*

On the basis of what is presently known, one can conjecture that even
if the pairing rule does not hold in the above form it could still hold
in the form of an inequality: $-\media{\s}_+\le\l^+_j+\l^-_j\le0$
({\it weak pairing rule}).  \*

\0{\it (E) Smoothness of the Lyapunov spectrum:} the Lyapunov exponents
become for large $N$ a smooth function of their index. This means that,
with the labeling of the exponents as in (D) above, if one draws a graph
of $x=\fra{j}D\to \l^+_j\=f_N(\fra{j}D)$, then in the ``thermodynamic
limit'' ($N\to\io$ with constant density for particle systems; in the
case of fluid systems the role of $N$ will be taken by the Reynolds
number) $f_N(x) \tende{N\to\io} f_\io(x)$ where $f_\io(x)$ is a smooth
increasing function of $x\in[0,1]$.\*

Evidence for the generality of such property comes from [LPR], and
[ECM1], [SEM], and quite likely it holds for all the models we consider
in \S3.
\*
We make the following remarks on the properties (A) to (E).

1) First we note that the irreversible entropy production $\media\s_+$
in (A) results in a phase space volume contraction. This implies in turn
that the attractor that we denote $A_0$ (which by property (C) is just
the full phase space $\CC$) will contain an invariant set $A$ of zero
Liouville measure and dimension equal to the {\it fractal dimension of
the motions} (and strictly less than that of the phase space, see below)
{\it but of probability $1$} with respect to the statistics of the
motions generated by the dynamics $S$ from initial data chosen randomly
with respect to the Liouville distribution $\m_0$.  \\
It is convenient, therefore, to distinguish between the attractor $A_0$
as it is usually defined in the literature (which is a closed set for
virtually all adopted definitions) and our sets $A$. The latter are not
uniquely defined, but they are in an obvious sense more intrinsically
related to the motions. It can very well be that $A_0$ is smooth and
even coincides with the full phase space, as is the case when (C) holds,
while $A$ is much smaller (and is a {\it fractal}). Thus in this paper,
unlike in most established conventions, {\it we shall call $A$ the
attractor}: it will not matter which particular $A$ one considers.\\ We
adopt, as definition of the {\it fractal dimension of the motions} (\ie
of $A$), the Kaplan Yorke definition (also called the {\it Lyapunov
dimension}, [ER]) The latter is, probably, [ER], quite generally equal to the
Hausdorff dimension of those sets $A$ which have the smallest Hausdorff
dimension and which are visited with frequency $1$ by almost all, with
respect to the Liouville distribution $\m_0$, motions [ER] p. 641.

2) The above properties (D,E) imply that the attractor $A$ for the
motions with a given energy is a {\it fractal set} with a dimension
close to the full dimension $2D$: the fractal dimension will be, in
fact, of the order of $2D-O(\media{\s}_+\l_{\max}^{-1})D$ as
immediately follows if one adopts, as above, the Kaplan Yorke definition
of fractal dimension.  Note that (E) and the above weaker pairing rule
are sufficient for this conclusion. {\it Systems for which smoothness
and the (weak) pairing rule hold do show dimension reduction}, \ie the
attractor in phase space has a dimension which is {\it macroscopically
different from that of the phase space itself}.

3) Reversibility implies an important property of the attractor $A$: if
$A$ is an attractor for the forward motion then $A_-=iA$ is an attractor
for the backwards motion and, more generally, the statistical
properties as $t\to\pm\io$, of the motions generated by initial data
randomly chosen with respect to the Liouville distribution $\m_0$ are
trivially related.

4) The basic properties for the validity of our results for
nonequilibrium stationary states are chaoticity (C) and reversibility
(B). If (C) holds the existence of the SRB distribution, \ie of a
probability distribution describing the asymptotic statistics of the
motions of a system evolving with a dynamics $S$, whose initial data are
chosen randomly with respect to the Liouville distribution $\m_0$ in the
phase space $\CC$, can be proved as a theorem:
\*

{\it Theorem: if a system $(\CC,S)$ is a
transitive Anosov system then it admits a SRB distribution, [S2].}  \*
\*

In conservative systems in equilibrium, verifying (C), the distribution
$\m$ is the same as the Liouville distribution $\m_0$ itself (which is
invariant by the Liouville theorem): see [S1,S2],[AA]. Hence (C) implies
the ergodic hypothesis in this case and the attractor $A$ can be taken
to be the full phase space $\CC$.

In this paper we are interested in dissipative systems satisfying (A)
where new phenomena occur and $\m$ {\it is not} the Liouville
distribution $\m_0$. For systems verifying (A),(B),(C) we prove a {\it
fluctuation theorem}, see \S7, which is our main technical result.
\*
\0{\it\S3 Models.}\numsec=3\numfor=1 \*

We now list a number of models to which our theory can conceivably be
applied. All these models contain thermostat mechanisms in order to
enable the systems to reach a non equilibrium stationary state in the
presence of an imposed external field. Model 1 is a model related to
electrical conductivity, models 2 and 3 are related to shear flow, model
4 to heat conduction and model 5 to a fluid mechanics model for
turbulent flow.

We distinguish, as in \S2, between the phase space $\FF$ over which the
system evolves according to the equations of motion and the collision
phase space $\CC$ consisting of the timing events (``Poincar\'e section
of $\FF$'').

{\it The details of the models described here {\sl will not} be used in
the following since our main point is the generality of the derivation
of a fluctuation formula from the chaotic hypothesis and its
(ensuing) model independence. However, we include them for concreteness
and reference.}  \*

\0{\it Model 1}: a gas of $N$ identical particles with mass $m$,
interacting via a stable short range spherically symmetric pair
potential $\f$ and with an external potential $\f^e\ne0$, enclosed in a
box $[-\fra12L,\fra12L]^2$ and subject to periodic boundary conditions
and a horizontal constant external field $E \V i$ ($\V i$ is a unit
vector in the $x$--direction). The external potential will be just a
hard core interaction which excludes access to a number of obstacles
(hard disks, to fix the ideas) so situated that {\it every} trajectory
must suffer collisions with them.  The system is in contact with a
"thermostat" adding (or subtracting) energy so that the total internal
energy stays rigorously constant.  The equations of motion are:

$$\eqalign{
&\dot{\V q_j}=\fra1m \V p_j,\qquad \dot{\V p_j}=\V F_j+E\V i-\a(\V
p)\V p_j\cr &\V F_j\=-\sum_{i\ne j}\Dpr_{\V q_j}\f(\V q_j-\V
q_i)-\Dpr_{\V q_j}\f^e(\V q_j)\cr}\Eq(3.1)$$

\0with $j=1,\ldots,N$; $\a(\V p)=E\,\V i\cdot\sum_j \V p_j/\big(\sum_j
\V p_j^2\big)$ and $\V F_j$ is the force acting on particle $j$. The
$\a$--term incorporates the coupling to a "gaussian thermostat" and
follows from Gauss' ``principle of least constraint". The constraint
here is the constancy of the internal energy:
$$H(\V p,\V q)=\sum_{i=1}^N\fra{\V p_j^2}{2m}+\sum_{i<j}\f(\V q_i-\V
q_j)+\sum_i\f^e(\V q_i)\Eq(3.2)$$

\0a typical nonholonomic constraint; it follows then from Gauss' principle
that the force corresponding to the constraint is proportional to the
gradient with respect to $\V p_j$ of $H$. This model has been studied in
great detail, in [CELS], in the case $N=1$ and $\f=0$ and $\f^e$ a hard
core potential as above, making it a Lorentz model for electrical
conductivity if $E$ is an electric field; a similar model has been
investigated numerically in [BEC].  It is part of a wide class of
models, together with the following models 2,3, whose interest for the
theory of non equilibrium stationary states was pointed out in [HHP],
[PH], where one can find the first studies performed in the context in
which we are interested. The dimension of the phase space $\FF$ of this
system is $d_0=4N-1$ and that of $\CC$ is $d_0-1=2D$ with $D=2N-1$.
{\it The phase space ``contraction'' rate, \ie the divergence of the
right hand side of \equ(3.1), is $D\s(x)=D\a(x)$}, which can be written
in the form:

$$D\s(x)=D\a(x)=D \fra{\e(x)}{D kT(x)}\Eq(3.3)$$

\0where $\e(x)$ is the work done on the system per unit time by the
external field and $k T(x)$ is $\fra1D\sum_j\fra{\V p_j^2}m$ which, if $k$
is Boltzmann's constant, defines a {\it kind of kinetic temperature}:
hence the name of {\it entropy production rate} per (kinetic) degree of
freedom that will be occasionally be given to
$\s(x)$.\footnote{${}^2$}{\nota If $\bar p$ is the average of $\V
i\cdot\V p_j$, then $\fra1N\sum_j \media{\V p_j^2}= \bar p^2+m
k T$ and we see that $T(x)$ cannot be identified with the temperature
unless we neglect $\bar p^2$ compared to $mk T$, \ie we identify the
peculiar momentum, needed for the proper definition of the temperature
with the ordinary momentum (which should not be done at large $E$ where
one cannot identify $\fra12\media{\sum_j \V p_j^2}$ with $m kT$,
with $T$ being the usual temperature).}  Note that $\s(x)$ does
not have a definite sign.

It has been proved, [CELS], that for $N=1$ and small $E>0$ the average
$\media\s_+$ is positive, \ie the system is dissipative in the sense of
\S2. There seems to be no reason to think that $\media\s_+$ is not
positive. For the above model with $N>1$ no experiments are available
yet on the pairing rule or the Lyapunov spectrum
smoothness. Nevertheless one can present an argument for the validity of
the pairing rule, which may sound convincing but that we have been
unable to substantiate mathematically.\footnote{${}^3$}{\nota Suppose
that the equation \equ(3.1) is modified into the same equation with
$\a(\V p)$ constant. Then [Dr],[ECM1] prove that the $4N$ Lyapunov
exponents can be paired so that the sum of the corresponding pairs is
just $-\a$. In model 1 $\a$ is {\it not constant}: however it has an
average value $\media{\a}$ which is constant on the attractor, with
probability $1$ with respect to the choice of the initial data (with
distribution $\m_0$): therefore we may hope that ``things go as if''
$\a$ was constant: hence the Lyapunov exponents should be so paired that
their sum is $-\media\a$. {\it This is not yet the above full pairing
rule} because there we assert in addition that half of the exponents are
positive and half are negative: and this only ``follows'' if
reversibility (B) is also used.} In this case the time reversal map $i$
is just the usual $i:(\V q,\V p)\to (\V q,-\V p)$.  \*

\0{\it Model 2}: a shear flow in a two dimensional container $[-\fra12
L,\fra12 L]^2$ where the particles evolve on a moving background running
with velocity $\V i\g \tilde y$ in the $x$ direction proportional to the
height $\tilde y$ in the $y$ direction, is measured with respect
to that of the center of mass of the particles. The background exercises
a drag on the $j$--th particle located at height $y_j$ proportional to
its peculiar velocity: $\dot{\V q}_j-\V i \g\tilde y_j$ with respect to
the background. The introduction of $\tilde y_j$ instead of the usual
$y_j$ is due to the boundary conditions we choose (see below), which do
not keep the height of the center of mass of the particles fixed. For
large $N$ the difference between $y_j$ and $\tilde y_j$ will become
negligible (see comment 6) in \S8). Similarly for large $N$ (and large
$L$ with $n=NL^{-2}$ fixed), the forcing $\g$ should be the shear rate
in the fluid, \ie $\g=\fra{\dpr u_x}{\dpr y}$, where $u_x(y)=\V i\g y$
is the average (local) velocity of the particles in the fluid at height
$y$, as is indeed found in the computer experiments for this system,
[ECM2].  If $\V F_j$ is as in the second equation of \equ(3.1) and
$\f^e=0$ and $\V q_j\=(x_j,y_j)$, the equations of motion are, using
Gauss' principle of least constraint to keep the internal energy fixed:
$$\eqalign{ &\fra{d^2}{dt^2}\V q_j=\fra1m\V F_j-(\dot{\V q}_j-\V i\g \tilde
y_j)\,\a\cr &\txt \a(x)\= \fra{m\g}{\sum_j \V
p_j^2}\Big(\sum_{j=1}^N\fra{p_{jx}p_{jy}}m -\fra12\sum_{i\ne j}
F_{x\,j,i}\,(y_i-y_j)\Big)\cr}\Eq(3.4)$$

\0with $j=1,\ldots,N$. Here $\V p_j=m(\V {{\dot q}}_j-\V i\g \tilde
y_j)$ is the peculiar momentum relative to the background flow; $\V
F_{j,i}$ is the force on particle $j$ due to particle $i$ and $\a$ is
again defined so that \equ(3.2) is a constant of the motion; finally
$\g$ plays here the role of a forcing field as $E$ did in model 1.  One
imposes periodic boundary conditions on the horizontal direction; on the
vertical direction a natural boundary condition is perfect reflection
against the walls at $y=\pm\fra12 L$.  This model has been extensively
studied, numerically, in [ECM1], [ECM2] with somewhat different boundary
conditions.

One can suppose that the total horizontal component of the peculiar
momentum and the horizontal position of the center of mass denoted,
respectively, $P_x, X_x$, are $0$: this is consistent with the equations
of motion. We shall refer to $P_x$ and $X_x$ as {\it conserved
quantities}: but one should bear in mind that they are such in an
"improper" way because they are conserved only if their initial values
are $0$: $P_x$ in fact relaxes to $0$ with a Lyapunov exponent which in
general is not zero (and equals the time average $\media{\a}_+$ of $\a$).

If it is assumed that $P_x=X_x=0$ and if one recalls that also $H$ is a
constant of the motion and one imposes \ap its value, then the dimension of
the phase space $\FF$ is $d_0=4N-3$, which we write $d_0=2D+1$ for
uniformity with the notation in model 1, so that $D=2N-2$. The phase
space contraction rate, \ie the divergence of the r.h.s. of the equation
of motion regarded as first order equations for $\V p,\V q$ is $D\s(x)$
with:

$$\s(x)=\a(x)+\g\fra{\sum_j p_{xj}p_{yj}}{D \sum_j\V p_j^2}=\a(x)+\g
O(N^{-1})=\fra{\e(x)}{D k T(x)}\Eq(3.5)$$

\0where $T(x)$ can {\it actually} be interpreted as a kinetic
temperature, so that $\s(x)$ can also be called the entropy production
rate.

Numerical experiments with $N$ up to $864$ show that $\media{\s}_+>0$,
[ECM1], ECM2], [SEM]. Such papers also provide (strong) evidence for the
pairing rule and some (weak) evidence for the smoothness.
The time reversal map in this case is {\it not} the usual velocity
reversal but $i:(x,y,p_x.p_y)\to (x,-y,-p_x,p_y)$.

\* \0{\it Model 3}: this is a model for a shear flow, produced by
boundary forces, in contrast to model 2 where the shear is produced by
body shear forces.

The flow proceeds in a two--dimensional container $[-\fra{L}2,\fra{L}2]$
and the equations of motion are simply:

$$\dot{\V q}_j=\fra1m \V p_j,\qquad \dot{\V p}_j=\V F_j\Eq(3.6)$$

\0supplemented by periodic boundary conditions on the horizontal
direction and shear generating boundary conditions in the vertical
direction:
$$\o'=f(\o)\Eq(3.7)$$

\0where $\o$ is the collision angle formed by the incoming velocity with
the $x$-axis, counted counterclockwise for collisions at $y=L$ and
clockwise for collisions at $y=0$; $\o'$ is the outgoing velocity
angle formed with the $x$-axis, counted clockwise at $y=L$ and
counterclockwise at $y=0$.

With the above angular conventions, $\o'=\o$ represents the ordinary
elastic collision.  We shall consider a "shearing collision rule"
$\o'=f(\o)$, $\o'\le\o$, where $f$ is a reversible collision rule.
Reversibility here has the literal meaning: the collision
obtained by reversing the particle velocity after a given collision
(\ie the incoming collision with an angle $\p-f(\o)$) produces
afterwards the reverse of the original collision, (\ie $\p-\o$).  This
is:
$$\p-\o=f(\p-f(\o)),\qquad f(\o)\le\o\Eq(3.8)$$
where the first condition is the {\it reversibility} condition and the
second is the {\it shearing} condition. The equations \equ(3.8) can be
solved by simply thinking that the graph of $f(\o)$ as a curve
$\o\to(\o,f(\o))\in[0,\p]^2$ is a concave arc connecting the point
$(0,0)$ with the point $(\p,\p)$, symmetric by reflection around the
secondary diagonal of $[0,\p]^2$. Furthermore one imposes that the
collisions preserve also the horizontal total momentum and the total
energy. Since the horizontal momentum of the colliding particle (say
particle $1$) changes (by $|\V p_1|(\cos\o'-\cos\o)$) one can impose the
two conservation laws by gaussian minimal constraints, \ie by requiring
that the variation of the other momenta is:

$$\V p'_j=(1+\vartheta)\V p_j+\beta \V i\Eq(3.9)$$

\0for suitable multipliers $\th$ and $\b$, which after a brief
calculation leads to explicit expressions for $\b,\th$, with
$\b=O(N^{-1})$ while $\th=O(N^{-2})$, so that the relations \equ(3.9)
generate corrections which, while enforcing the constraints, can be
regarded as negligible for large $N$ (see comment 6) in \S8).

The symmetry of $f$ also guarantees that the collision rule \equ(3.8)
corresponds to a gaussian constraint\annota{4}{For this purpose one has
to think, [CL], that the collision rule \equ(3.8) is realized as a limit
of a gaussian constraint rule acting on a tiny corridor of width $\d$
expanding vertically the container at the top and bottom, where the
particles can enter from the inside but are subject to a horizontal
field $\pm E_\d$ (the sign depending on whether the particles are in the
upper or lower corridor): the field will produce a bias in the
scattering angle such that the particle will come out with an angle
different from the incoming angle. The constraint is that the kinetic
energy of the particles inside the corridors does not change; the
particles colliding with the external corridors walls are just perfectly
reflected. In the limit as $\d\to0$ and $E_\d\to\io$ (at suitable rates)
a reflection rule like \equ(3.8) is realized with a special $f$. By
letting $E_\d$ depend on the distance to the corridor boundaries
essentially any $f$ can be realized, in the limit $\d\to0$.}  (forcing
$|\V p_1|$ to stay constant in the collision, while changing the
direction with respect to the elastic collision), [CL].

For concreteness one can take, following [CL], $f$ such that the above
arc is an arc of a circle centered on the secondary diagonal and passing
through the indicated points. The circle curvature will be a measure of
the shear strength. The dimension of the phase space $\FF$ of this
system is $d_0=4N-3$, if we fix the energy, the horizontal total
momentum and the horizontal position of the center of mass.  We write
$d_0=2D+1$ as in the previous models so that the dimension of the
collision space $\CC$ is $2D$ with $D=2N-2$. We also suppose, naturally,
that the collisions with the walls are among timing events. Then at
every collision there is a reduction of phase space volume
$\fra{\sin\o'\, d\o'}{\sin\o\, d\o}=\fra{\sin f(\o)}{\sin\o}\, f'(\o)$,
[CL]. If we define $n(x)=1$ when $x$ is a collision with the wall and
$n(x)=0$ otherwise, the phase space contraction can be conveniently
written as $e^{-D\t(x)\s(x)}$ with $\t(x)$ equal to the time elapsing
between the collision at $x$ and the next at $Sx$ and with the entropy
production rate $\s(x)$ defined by:

$$ D \s(x)=\fra{n(x)}{\t(x)}\,\log (1-\fra{\sin
f(\o)}{\sin\o}f'(\o))^{-1} \Eq(3.10)$$

\0This model has been studied in detail in [CL].\footnote{${}^5$}{\nota It
is not difficult to see, by thinking of the constraint as a (limit of)
gaussian constraints as above, to see that {\it also} in this model
$D\s(x)$ is an ``entropy production rate''.}

There is numerical evidence that if $f(\o)\ne\o$ then $\media{\s}_+>0$. The
pairing and smoothness properties have not yet been studied. The time
reversal operation is the ``usual'' one (see model 1).

Models 1 and 2 are easier to interpret as dynamical systems than this
model, but they are physically somewhat artificial, in that the gaussian
thermostat is a rather unconventional model for a thermostat and the
shear force is a body force rather than the usual boundary force. In
this respect model 3 is better as its unphysical reflection laws are a
boundary effect which only produces the net effect of generating a shear
on the system.  \*

{\it Model 4)} is a model for heat conduction considered in [HHP], [PH].
In a box, $[-L-H,L+H]\times[-\fra{L}2,\fra{L}2]$, $N$ particles move
interacting via a short range pair potential; the boundary conditions
are perfect reflection horizontally and periodic vertically; the
particles are subject to the nonholonomic constraint that the total
kinetic energy in the left part of the box
$[-L-H,-H]\times[-\fra{L}2,\fra{L}2]$ and in the right part of the box
$[H,L+H]\times[-\fra{L}2,\fra{L}2]$ have constant values, denoted,
respectively by $\fra{L}{2L+2H}N k T_-$ and $\fra{L}{2L+2H} k T_+$,
$T_+\ge T_-$, \ie obey:

$$\F_\pm=\sum_{j=1}^N\chi_\pm(x_j)\fra{{\V
p_j}^2}{2m}=\fra{N}2\fra{L}{H+L} kT_\pm\Eq(3.11)$$

\0Here $\chi_\pm $ are the characteristic functions of the left and
right parts of the box which have to be interpreted as {\it plates} of
thickness $L$ {\it at temperatures $T_+$ and $T_-$} respectively; $\V
q\=(x,y)$. In addition we impose that the total energy (given by
\equ(3.2) with $\f^e=0$) of the gas $\F_0=H$ is exactly conserved.

The constraints are implemented using Gauss' principle, \ie by a force
proportional to the gradients with respect to the $\V p_j$'s of $\F_\pm$
and $\F_0$ (which are simply $\chi_\pm(x_j)\fra{\V p_j}{m}$,
$\fra{\V p_j}{m}$, respectively), leading to the equations of motion:

$$
\dot{\V q}_j=\fra1m {\V p_j},\qquad \dot{\V p}_j={\V F}_j-\a_+
\chi_+(x_j)\V p_j-\a_-\chi_-(x_j)\V p_j-\a_0 \V p_j\Eq(3.12)$$

\0with $\a_\pm,\a_0$ are defined so that $\F_\pm,\F_0$ are exact
constants of motion. The values of $\a_\pm,\a_0$ can be easily computed;
their general expression will not be needed here. For the
purpose of illustrating once more that the resulting forces will lead to
a reversible dynamics we give their expression in the simple case in
which only $\F_\pm$ are imposed: in this case the values of $\a_\pm$ are
relatively simple and $\a_0$ is not present. One finds:

$$\a_\pm=\fra{
\sum_j \big(\fra{{\V p}_j}m
\cdot\V\dpr \chi_\pm(x_j)\fra{\V
p^2_j}{2m}+\V F_j\cdot\fra{\V p_j}{m}
\chi_\pm(x_j)\big)}{ \sum_j \chi_\pm(x_j)^2{\V p_j^2} }\Eq(3.13)$$

Going back to \equ(3.12) we note that, with the three mentioned
constraints, model 4 should be a quite realistic model for heat
conduction.  The dimension of the phase space $\FF$ is $d_0=4N-3$, which
again we write as $d_0=2D+1$, with $D=2N-2$, so that the dimension of
the collision space $\CC$ is again $2D$. The phase space contraction
rate is in this case:

$$D\s(x)=\a_+(x) 2N_++\a_-(x) 2N_-+\a_0(x)2N+O(N^{-1})\Eq(3.14)$$

\0if $N_\pm$ denote the number of particles in the right and left
``plates''. Eq. \equ(3.14) could also be interpreted as in the previous
models as an entropy production rate.  Some numerical evidence that
$\media{\s}_+>0$ if $T_+> T_-$ can be found in [PH],[HHP]. No evidence
for pairing or smoothness rules seems available. The time reversal map
is the ``usual'' one, see model 1.

The above model equations can be made smoother by replacing $\chi$ by
a smoothed version of the characteristic functions of the plates; say by
functions which are $\=1$ except within a distance of the order of the
interaction range from the inner boundaries of the plates: in this
region $\chi_\pm$ decrease gently from $1$ to zero.  \*

{\it Model 5)}: this is a model related to turbulent flow,
obtained from the Navier Stokes (NS) equations. We consider the NS
equations in a box $[-\fra{L}2,\fra{L}2]^3$, with periodic boundary
conditions and for an incompressible fluid. If the velocity field is
written in a Fourier series as:

$$\V u(\V x)=\sum_{\V k\ne \V0} e^{i\V k\cdot\V x}\V\g_{\V k}\Eq(3.15)$$

\0with $\V\g_\kk$ complex vectors with $\V\g_\kk=\lis{\V\g}_{-\kk}$
(reality of the velocity field) and $\g_{\,\kk}\perp \kk$
(incompressibility) then the NS equations become:

$$\dot{\V \g}_\kk=-i\sum_{\V k_x+\V k_2=\V k}(\V \g_{\V k_1}\cdot\V
k_2)\, \Pi_{\kk}\, \V \g_{\,\V k_1}+ R\V f_\kk-\n\kk^2\V
\g_{\,\kk}\Eq(3.16)$$

\0$\Pi_\kk$ is the orthogonal projection over the plane orthogonal
to $\kk$; $\n$ is the kinematic viscosity and $R\V f_\kk$ is the forcing
(of course orthogonal to $\kk$) which will be taken to be non zero only
for a few components with small $\kk$. Since $\kk=\fra{2\p}L \V n$ with
$\V n$ integer this means that the force acts only on the high length
scale components. For simplicity we may think that the forcing has only
two non vanishing components $R\V f_{\kk^0_1},R\V f_{\kk^0_2}$,
corresponding to two linearly independent wave numbers
$\kk^0_1,\kk^0_2$.\footnote{${}^6$}{\nota The simpler case of only one
non zero component can be trivial (\eg if the forcing acts on the
smallest $\kk$, $|\kk|=k_0$) and is therefore discarded here in favor of
the next to the simplest, [Ma].} The number $R$ therefore determines the
forcing strength and will be identified with the Reynolds number (we
keep the container size $L$ and the viscosity $\n$ fixed). We take
$\V\g_{\,\V0}\=\V0$ since it is the conserved center of mass velocity.

In order to obtain equations in the framework of this paper from
the phenomenological theory of Kolmogorov--Obuchov, [LL], we
shall assume that the above equations can be replaced by the following
simpler ones:

$$\eqalign{ \dot{\V \g}_\kk=&-i\sum_{\V k_1+\V k_2=\V k} (\V \g_{\V
k_1}\cdot\V k_2)\,\Pi_{\kk} \g_{\V k_1}+ \V f_\kk\kern1.cm\qquad
|\kk|<k_R\cr \dot\V \g_\kk=&-\a \V\g_\kk -i\sum_{\V k_1+\V k_2=\V k} (\V
\g_{\V k_1}\cdot\V k_2)\,\Pi_{\kk} \g_{\V k_1} \kern1.cm \qquad k_R\le
|\kk|<k_R +\n^{-1/2} \cr}\Eq(3.17)$$

\0Here, if $k_0=\fra{2\p}L$, the wave vector $k_R$ is the Kolmogorov
momentum scale $k_R=k_0 R^{3/4}$, ([LL] p. 122, (32.6)), so that if
$N_R$ is the number of wave vectors (``modes'') $\kk$ such that when
$k_0\le |\kk|\le k_R+\n^{-1/2}$ then $N_R\approx
(\fra{k_R}{k_0})^3\approx R^{9/4}$; then the phase space $\CC$ has
dimension $2N_R-2=2D$ with $D\approx R^{9/4}$, while $\FF$ has dimension
$d_0=2N_R-1$.\footnote{${}^7$}{\nota Taking into account the reality
and incompressibility conditions, forces the $\V\g_\kk$ to
have only two linearly independent components.}

This means that the equations for the amplitudes $\V\g_\kk$
corresponding to $\kk$'s in the {\it inertial range}, $k_0\le|\kk|\le
k_R$, are ``governed'' by the reversible Euler equations.  In the {\it
viscous range}, $|\kk|>k_R$ the dissipation phenomena will be idealized
by saying that the equations are simply such that only the modes $\kk$ with
$k_R<|\kk|< k_R+\n^{-1/2}$ have a non zero amplitude and
evolve in such a way as to keep the total energy constant.  This means
that the parameter $\a$ is an effective thermostat (or viscosity), which
has to be chosen so that the total energy is constant, \ie so that
$\fra{d}{dt}\sum_\kk |\V\g_\kk|^2=0$:

$$ \a(x)=\fra{\sum_\kk\V f_\kk\cdot\V \g_{-\kk}}
{\sum_{|\kk|>k_R}|\V\g_\kk|^2}\,{\buildrel def \over
=}\,\fra{\e(x)}{D_vkT(x)} \Eq(3.18)$$

\0Here $\e(x)$ and $D_vkT(x)$ are simply the numerator and denominator
of the fraction defining $\a(x)$, if $2N_{vR}$ is the number of modes in
the viscous range and one defines $2D_v=2N_{vR}$.

The Kolmogorov length $k_R^{-1}$ is introduced here phenomenologically
and we do not attempt at a fundamental derivation of \equ(3.17),
\equ(3.18). Therefore \equ(3.17) has to be regarded as a
phenomenological equation.

Note that $\a$ is proportional to the work $\e(x)$ per unit time and per
viscous degree of freedom performed on the system, which is dissipated
into heat, in order to keep the total energy constant: the
proportionality constant is $2D_v k T(x)$ with
$kT(x)\=\fra1{2D_v}\sum_{|\kk|<k_R}|\V\g_\kk|^2$ (which, however, is
{\it not} a constant of motion for \equ(3.17),\equ(3.18) because of the
imposed constraint that $\sum_\kk |\V\g_\kk|^2$ rather than
$\sum_{|\kk|>k_R} |\V\g_\kk|^2$ is constant). The phase space
contraction rate is in this case:

$$D_{v} \s(x)=D_v\a(x)=D_v \fra{\e(x)}{D_vkT(x)}\Eq(3.19)$$

\0Hence $D_v\media{\s}_+$ can be thought of as the average amount of
{\it energy dissipation} per unit time by the flow divided by the {\it
kinetic energy} contained in the {\it viscous modes}. The first quantity
plays a major role in Kolmogorov's theory, see [LL] p.119, and its
average is usually called $\e$, (see [LL], (31.1)). Since the kinetic
energy contained in the viscous modes can be thought of as a kind of
``temperature'' we see that $2D_v{\media\s}_+$, is proportional to the
entropy ``production rate''. More appropriately we can say that, for $R$
large, $2D_v{\media\s}_+$ is proportional, once more, to the ``energy
dissipation rate'' over a kinetic quantity equal to the average kinetic
energy contained in the viscous modes {\it if}, for large $R$, the two
quantities can be regarded as independent random variables.

Note, however, that for the above model \equ(3.17) (introduced, we believe,
for the first time here) there is no evidence for ${\media\s}_+>0$ or
for the pairing and smoothness rules. The time reversal map is simply
$i:\{\V\g_\kk\}\to\{-\V\g_\kk\}$.

\vskip1cm

\0{\it\S4 The introduction of the SRB distribution.}
\* \numsec=4\numfor=1

We now present a heuristic argument providing, in our opinion, a useful
characterization of the SRB distribution: this point of view is
important for the applications of \S7. Our purpose is to look at it from
a somewhat different perspective than usual and to show that it leads to
a new interpretation of the ergodic hypothesis and to a unification of
equilibrium and nonequilibrium statistical mechanics.

We deal with systems of $N$ particles verifying the properties (A,B,C)
of \S2 and we observe their motions $x\to S^nx$ at discrete times in the
collision space $\CC$ of dimension $2D$, see \S2.  It will be
very useful to keep in mind the paradigm of hyperbolic systems:
namely the Anosov map of the 2-dimensional torus $T^2$:
$$S\,\pmatrix{\f_1\cr\f_2}=\pmatrix{1&1\cr1&2\cr}\pmatrix{\f_1\cr\f_2}
\ \mod2\p\Eq(4.1)$$
which plays a role analogous to that of harmonic oscillators in
classical mechanics. This example is not only enlightening, but it is
really the main source of intuition. Note that this is a reversible map
if $i$ is defined as $i: (\f_1,\f_2)\to(\f_2,-\f_1)$, because
$iS\=S^{-1}i$.

Let $O$ be a fixed point on the attractor $A$ and let $W^u_O$ be the
unstable manifold $W^u_O$ of $O$ (dense on $A$, see (C),
\S2).\footnote{${}^8$}{\nota There might be {\it no such point} fixed point
$O$.  However a periodic orbit starting at a point $P$ and with period
$n$ would be a fixed point for $S^n$ and we could get all the following
conclusions by replacing $S$ with $S^n$, since the statistics of $S^n$
and that of $S$ coincide, when $S$ is chaotic enough. Thus assuming the
existence of a fixed point is not restrictive.}  The dimension of
$W^u_O$ is $D$, half that of the phase space $\CC$, see \S2.

For simplicity we shall suppose that $O$ is a time reversal invariant
fixed point $O=iO$; this assumption could be easily
relaxed.\footnote{${}^9$}{\nota It is not difficult to realize that in
the models 1,2,3,4 there are always periodic orbits which are time reversal
invariant, \ie such that $iO$ is also on the orbit; at least if one is
willing to limit the particle density in some interval (whose size may
depend on the range of the interaction). Also for model 5 it is very
likely that periodic motions (unstable, of course) do exist. Note that
since we are assuming (C), \S2, it is automatically true that there are
periodic orbits (\ie chaotic systems always have a lot of periodic
unstable orbits), [S2].}

The key idea on which we base our analysis is that the attractor $A$
should be considered to consist of the {\it smooth} $D$--dimensional
unstable manifold $W^u_O$ of $O$ (or of any other fixed point or
periodic orbit in $A$ with dense stable and unstable manifolds).  Of
course the manifold $W^u_O$ can only fill $A$ densely: we "lose" the
accumulation points. But all the information needed to perform time
averages should be already contained in $W^u_O$ itself, as we are only
interested in the averages of rather regular observables (\eg piecewise
smooth). We think that $W^u_O$ coincides with $A$ in the same sense in
which the rationals can be regarded as coinciding with the reals in
integration theory (which works only if one considers integrals of
smooth functions) and can be used to compute numerically the integrals
of smooth functions.  In the same way statistical averages with the
distribution \equ(2.1) should be computable by simply approximating them
with integrals over finite parts of $W^u_O$, like the sets $S^T\D$
obtained by ``blowing up'' with a large time iterate $S^T$ a small
connected surface element $\D$ of $W^u_O$ containing $O$.

In other words we want to regard the possible fractality of $A$ as a
rather irrelevant accident. We want to think of $A$ as {\it
essentially} identical to $W^u_O$: the latter surface folds over and
over again, being enclosed in the {\it bounded} phase space $\CC$. It
therefore folds itself in $\CC$ just as an uncut folio is folded into a
book, thereby generating an almost three dimensional fractal set out of
a two dimensional smooth manifold. But thinking of $A$ as an unfolded
manifold of half the space dimension ($D$ in our notation) leads to a
change in the usual point of view, which regards $A$ as a fractal set
with dimension close to $2D$.

Introducing forcing and friction (\ie passing from an equilibrium to a
stationary non equilibrium problem) should then {\it not} be thought of as
a real ``discontinuity'': which would be the case if one took
the viewpoint that one is passing from a nice smooth $2D$ dimensional
attractor $A=\CC$ to a nasty, strange, fractal $A\subset \CC$ with
dimension $2D- O(\a\l^{-1}_{\max})D$, {\it macroscopically different}
from $2D$ (as implied by the pairing (D) together with the smoothness
(E), \S2).

Rather it should be viewed as an insignificant deformation of the
unstable manifold $W^u_O$ which will fold itself in $\CC$ not
{\it exactly} as in the conservative case, but leave a few holes
between the "pages" to account for its global fractality.  This is a
change with respect to the conventional point of view for the case of
conservative systems: these are no longer really different from the
dissipative ones. Their attractor has, in the new ({\it unconventional})
sense, {\it exactly} half the dimension of the full phase space
(the dimension the conventional point of view attributed to them
is that of the {\it full} phase space, \ie twice as large).

The main consequence of such a viewpoint, besides the mentioned
unification of conservative and dissipative dynamics, is that it allows
us to think of the attractor as "unfoldable', with the consequence that
our intuition about the motion on the attractor is greatly enhanced.

This unfolded attractor, imagined as a flat infinite surface, attracts
exponentially fast nearby points: the approach to the attractor follows
the stable manifolds associated with the attractor points, which can be
thought of as {\it needles} sticking out of the attractor itself.  The
motion essentially consists, therefore, of an expanding (\ie as unstable
as possible) motion on the unstable manifold $W^u_O$.

We can now easily understand the statistics $\m$, \equ(2.1), on the
attractor, \ie the SRB statistics, as follows.

Let $U$ be a sphere with small radius $h$, centered at the fixed point
$O$; and let us ask how to compute the time average of an observable
$F$, if the initial data are chosen in $U$ with uniform distribution,
say, with a distribution absolutely continuous with respect to the
Liouville distribution.

Clearly the average of $F$ over a large time $T$ will be computable by
looking at the image $S^TU$ under $S^T$ for large $T$ and by imagining
$S^T U$ covered by the density into which the initial uniform density in
$U$ evolves in $T$ time steps. If we call $\D$ {\it the connected part}
of $W^u_O\cap U$ {\it this also means that we can regard the $S^T$
image, $S^T\D$, of {\it the connected part} of $W^u_O\cap U$ as a good
{\sl finite} approximation $S^T\D$ to our attractor}.

The set $S^TU$ will be extremely thin and it will ``{\it coat}''
the extremely large portion of $W^u_O$ defined by $S^T\D$, (\ie by our
``good finite approximation'' of the attractor), if we regard the
attractor as unfolded.

Let $dx$ be a surface element on $W^u_O$ and let us regard $S$ as a map
of $W^u_O$ into itself.  {\it We shall call $\L_u(x)$ the absolute value
of the jacobian determinant $ \dpr_u S(x)$ of $S$, as a map of $W^u_x$
into itself, at the point $x$}. In this way $\L_u(x)$ will be the
absolute value of the determinant of a matrix with a dimension equal to
that of $W^u_x$, \ie $D$.  Then we are interested in computing the
integral:
$$\ig_{S^T U}\r_T(x)\, F(x)\, dx\Eq(4.2)$$
where $\r_T(x)dx$ is the amount of mass in the cylinder with base $dx$,
which is the image of the cylinder in $U$ with base $S^{-T} dx$ and
height equal to the height $h$ of the initial "cloud of data"
$U$. Denoting $d_s$ (resp. $d_u$) the dimension of the stable (unstable)
manifold of $O$ (which in our case are $d_s=d_u=D$), this means that
$\r_T(x)dx $ is proportional to $h^{d_s}|S^{-T}dx|$, where $|S^{-T}dx|$
is the surface area of $S^{-T}dx$.  By the above definition of the local
expansion rate $\L_u(x)$, one has then:
$$\r_T(x)dx=const\, \L_u^{-1}(S^{-T}x)\ldots
\L_u^{-1}(S^{-1}x)dx\,\tende{T\to\io}\,const
\,\prod_{j=-\io}^{-1}\L_u^{-1}(S^jx)\,dx\Eq(4.3)$$
which is, clearly, a formal relation because $\r_T$ tends to $0$ as
$T\to\io$.  Note, however, that \equ(4.3) implies that the {\it ratios}
between $\r_T(x)$ and $\r_T(x')$, with $x,x'$ in the surface elements
$dx,dx'$, {\it are well defined}, even in the limit as $T\to\io$.

The equation \equ(4.2) is for $T$ large already a ``good approximation''
for the SRB distribution. It shows that the statistical averages
should be computable by looking at a large part of $W^u_O$, namely at
the finite approximation of the attractor $A$ called $S^T \D$ above, and
by imagining it {\it coated} with a density $\r_T(x)$, and then using
\equ(4.2).

The existence of the limit as $T\to\io$ of \equ(4.2)
can be seen by remarking that the limit can in fact be
written as an integral over phase space, in spite of the fact that
$\r_T(x)$ tends manifestly to $0$ as $T\to\io$.  For, when $T\to\io$
what really matters is the amount of mass ending up inside a generic
little square $E$ in the phase space $\CC$, with center $x_E$.  Since $E$
will be cut many times by $S^T\D$ we can imagine that the various "pieces" of
$S^T\D$ intersecting $E$ are piled up "vertically" in $E$: the figure
below shows a picture for the simple case \equ(4.1).

\figini{c1gcpaper}
\8<%!PS-Giovanni-1.0>
\8<% figura 1 in coen.tex: 43 linee>
\8<%%EndComments>
\8</p {1 5 sqrt add 2 div} def /m {1 5 sqrt sub 2 div} def >
\8</normp {1 p 1 sub dup mul add sqrt} def>
\8</normm {1 m 1 sub dup mul add sqrt} def>
\8</vxp {1 normp div} def>
\8</vyp {p 1 sub normp div} def >
\8</vxm {1 normm div neg} def>
\8</vym {m 1 sub normm div neg} def >
\8<      /L1 {150} def     %lunghezza assi stabile/instabile>
\8<      /L0 {100} def     %lato>
\8<      /L {300} def      %periodo>
\8<       /N {10}  def     %numero di tratti>
\8</h {L0 vxp div} def     %altezza accettabile modulo L >
\8<>
\8</xyps {dup vxp mul exch vyp mul} def          % necessita di s>
\8</xyms {dup vxm mul exch vym mul} def          % necessita di s>
\8<>
\8</instab {vxp mul dup vyp mul exch vxp mul moveto L0 xyp lineto stroke} def>
\8<>
\8</mmod {div dup truncate sub} def  %necessita di a b>
\8<>
\8</punta0 {newpath >
\8<0 0 moveto -10 5 lineto -10 -5 lineto closepath} def>
\8</dirpunta {gsave 2 copy translate 3 -1 roll sub 3 1 roll >
\8<exch sub atan rotate punta0 fill grestore} def>
\8<>
\8<gsave>
\8<%100 400 translate>
\8<150 0 translate>
\8<     0 0 moveto L1 xyps lineto stroke>
\8<     0 0 L1 xyps dirpunta>
\8<     0 0 moveto L1 xyms lineto stroke>
\8<     0 0 L1 xyms dirpunta>
\8<     L0 xyms 2 copy moveto L0 xyps exch  3 1 roll  add 3 1 roll add exch>
\8<        lineto stroke>
\8<     L0 xyps 2 copy moveto L0 xyms exch 3 1 roll add 3 1 roll add exch>
\8<        lineto stroke>
\8</xydzdw { 4 2 roll 2 copy moveto exch add 3 1 roll add>
\8< exch lineto stroke} def>
\8</instab {vxp mul dup vyp mul neg exch vxp mul 2 copy >
\8<moveto L0 xyps  exch 3 1 roll add 3 1 roll add exch lineto stroke} def>
\8<>
\8<0 1 38 {L mul vyp mul L mmod L mul dup h lt {instab L} if pop} for>
\8<>
\8<grestore>
\figfin

\eqfig{260pt}{150pt}{
\ins{265pt}{62pt}{$\x$}
\ins{235pt}{77pt}{$E$}
\ins{60pt}{83pt}{$\h$}
}{c1gcpaper}{}
\*

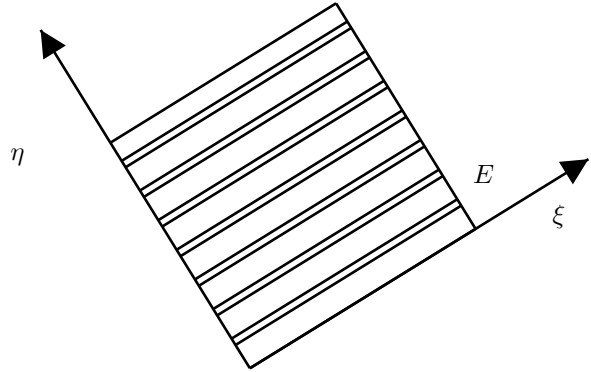
\0{\nota\it Fig. 1: The parallel lines are intersections of the finite
approximation for the attractor, $S^T\D$, with the set $E$, represented
by a square. The $\x,\h$ axes are ``parallel'' to the unstable and
stable manifolds $W^u_O,W^s_O$ respectively. Each of them is coated,
eventually, by the image $S^TU$ of $U$ which gives them a thickness,
(not shown). In the case of the map \equ(4.1) the parallel lines are
generated, if one moves on $W^u_O$ away from $O$, in the following
typical order: from bottom to top first one draws, successively, the
lower line of each pair; then one draws the second, then one should draw
a third series of lines above the second and keep going in this way
until the endpoints of $S^T\D$ are reached. For $T\to\io$ the parallel
lines fill densely $E$.\vfill} \*

\0If $\m_T(E)$ is the total mass initially in $U$ ending up in $E$
after time $T$, we can rewrite \equ(4.2) as:

$$\sum_E \m_T(E) F(x_E)\Eq(4.4)$$

\0provided $E$ is so small that we can neglect the variation of $F$
inside $E$, and that the $E$'s pave the phase space.  Suppose we set up
a coordinate system in the small box $E$ (which is a box with
full dimension $2D$) so that the ``horizontal'' coordinates are called $\x$
and the ``vertical'' ones $\h$. A point in $E\cap A$ is denoted $x(\x,\h)$
and the surfaces of constant $\h$ are connected surface elements of the
unstable foliation $W^u$ in $E$ (a foliation of a set $E$ is a family of
disjoint connected surfaces whose union is $E$), while those of constant
$\x$ are connected surface elements of the stable foliation $W^s$.

Then we see that \equ(4.2), before the limit as $T\to\io$ is taken, can
be expressed as a sum over the connected parts of the
surface\footnote{${}^{10}$}{\nota Which, we recall, represnts the finite
approximation to the attractor defined above.} $S^T\D$ that fall in $E$
(the parallel lines in the above figure)). If $\{\h_j\}$ representS the
$j$-th line then we can write \equ(4.2) as:
$$\sum_{\h_j}\ig_{\{\h_j\}}\r_T(\x,\h_j) F(x(\x,\h_j)) \,d\x\tende{T\to\io}
\ig_E\lis\r^\h(\x)du\,\n(d\h)\,F(x(\x,\h))\Eq(4.5)$$

\0where $d\x$ denotes an area element on $\{\h_j\}$. In the limit
$T\to\io$, while $\r_T(\x,\h_j)$ tends to $0$ the number of lines
$\{\h_j\}$ tends to infinity and the sum over the surface elements of
$S^T\D$ that cross $E$ {\it should converge} to an integral over $\h$
and $\x$ with respect to some measure $\lis\r^\h(\x)\n(d\h)d\x$ with
both the density $\lis\r^\h(\x)$ along the unstable manifold and the
measure $\n$ well defined. The measure $\n$ will give us the detailed
information on how the various pieces ({\it layers}, or lines in fig. 1)
of $W^u_O$ intersecting $E$ pile up and the distribution of the gaps
between them in $E$,\footnote{${}^{11}$}{\nota Note that ``gap'' here
does not mean an actually empty region: since $W^u_O$ is dense in $E$
(by (C), \S2) there can be no open regions in $E$ which are not crossed
by one connected part of $W^u_O$. In general, one should think of $\n$
as supported by a dense ``Cantor set''.} hence on its fractal nature; on
the other hand $\lis\r^\h(\x)$ will be a function such that the ratios
$\ig \lis\r^\h(\x)d\x/ \ig
\lis\r^{\h'}(\x)d\x$ should tell us the ratio of the masses of $U$
ending up near the pieces of unstable manifold passing through $\h$ and
$\h'$, inside $E$, which should be well defined, by \equ(4.3), in the
limit $T\to+\io$ as argued above.
\*\*

\0{\it\S5 The thermodynamic analogy.}  \numsec=5\numfor=1 \*

In this section we describe theoretical difficulties with the heuristic
analysis of \S4 and with a mathematical proof of the existence of the
limit \equ(4.2). The solution to the difficulties that will be pointed
out necessitates the introduction of more refined ideas and eventually
the use of Markov partitions. We first point out the difficulty.

If the analysis of the previous section is correct, \ie if
$\lis\r^\h(\x)$ really exists we should be able to ``calculate'' it, at
least formally. While it is evident that the function $\lis\r^\h(\x)$ is
defined up to a constant for each $\h$ such that $(\x,\h)\in W^u_O$,
(because of formula \equ(4.3) and the comment following it), it is much
less evident that $\lis\r^\h(\x)$ behaves reasonably regularly in
$(\x,\h)\in E\cap W^u_O$.

Clearly the right hand side of equation \equ(4.3) can be used to compare
the values of $\lis\r^\h(\x)$ and of $\lis\r^\h(\x')$ if $x=(\x,\h)$ and
$x'=(\x',\h)$ are points of $E\cap W^u_O$ with the same $\h$ (note that
in such case only few of the factors in \equ(4.3) have ratios really
diferent from $1$, because $S^{-j}x$ and $S^{-j}x'$ approach $O$
exponentially fast and start close on a connected part of
$W^u_O$). However if we compare $\lis\r^\h(\x)$ and $\lis\r^{\h'}(\x)$
with the same $\x$ and $(\x,\h),(\x,\h')\in E\cap W^u_O$ we run into the
difficulty that the distance between $(\x,\h),(\x,\h')$ {\it measured
along $W^u_O$ may be extremely long}, in fact as long as we please by
varying the two points on the surface $\x=constant$. Therefore the
function $\lis\r^\h(\x)$ {\it might vary quite irregularly in $E$}.
Hence we see that the existence of a limit in \equ(4.5) is not so
obvious even though \equ(4.3) provides immediately an expression for the
ratios of the limit density on the set $W^u_O\cap E$.

We {\it must} find an alternative way to control the variations of such
a function in $E\cap W^u_O$ in order to argue that $\lis\r^\h(\x)$ is
well defined. The number of connected components of $W^u_O$ in $E$ is
denumerable and we cannot expect that the SRB distribution is supported
by a denumerable set of $d_u$--dimensional surfaces. Hence we are in a
position similar to when attempting to define the integral of a
continuous function over a segment from the knowledge of the function at
the rational points on the segment: this is possible only if the
function is not too wildly changing from point to point.

The resolution of this difficulty proceeds in two steps. First we push
the analysis of the variability of $\lis\r^\h(\x)$ just given somewhat
further to arrive at the equations \equ(5.3), \equ(5.4) below which are
useful to illustrate the development of the thermodynamic analogy that
gave rise to the {\it thermodynamic formalism}.  This will enable us, in
\S6, to discuss the proper solution to the problem of the existence of
the limit \equ(4.5) and of the function $\lis\r^\h(\x)$, based on this
thermodynamic analogy.

Let $x=(\x,\h), x'=(\x,\h')$ be points of $E\cap W^u_O$ and let $d\x$
and $d\x'$ be two infinitesimal surface elements in $E\cap W^u_O$ at
different heights $\h'$ and $\h$, {\it corresponding} to each other, in
the sense that the stable manifolds through $d\x\subset W^u_x$ intersect
the unstable manifold $W^u_{x'}$ exactly on $d\x'$, see fig. 2
below. Then the masses $\lis\r^\h(\x)d\x$ on the segment $d\x$ and on
$\lis\r^{\h'}(\x)d\x'$ have a ratio that can be computed by using
\equ(4.3) and by remarking that the ratio of the areas $d\x/d\x'$ is:
$$\fra{d\x}{d\x'}=\fra{|S^{-1}(Sd\x)|}{|S^{-1}(Sd\x')|}=
\fra{|S^{-2}(S^2d\x)|}{|S^{-2}(S^2d\x')|}=\ldots=
\fra{|S^{-M}(S^Md\x)|}{|S^{-M}(S^Md\x')|} \Eq(5.1)$$
hence:
$$\fra{d\x}{d\x'}=\Big(\prod_{j=0}^{M-1}
\fra{\L_u^{-1}(S^j\x)}{\L_u^{-1}(S^j\x')}
\Big)\fra{|S^M d\x|}{|S^M d\x'|}\Eq(5.2)$$
See Fig. 2 for an illustration:

%\figini{c2}
\figini{c2gcpaper}
\8<%!PS-Giovanni-1.0>
\8<% figura 2 in coen.tex: 43 linee>
\8<%%EndComments>
\8</p {1 5 sqrt add 2 div} def /m {1 5 sqrt sub 2 div} def >
\8</normp {1 p 1 sub dup mul add sqrt} def>
\8</normm {1 m 1 sub dup mul add sqrt} def>
\8</vxp {1 normp div} def>
\8</vyp {p 1 sub normp div} def >
\8</vxm {1 normm div neg} def>
\8</vym {m 1 sub normm div neg} def >
\8< /L1 {150} def %lunghezza assi stabile/instabile>
\8< /L0 {100} def %lato>
\8< /L {300} def %periodo>
\8< /N {10} def %numero di tratti>
\8</h {L0 vxp div} def %altezza accettabile modulo L >
\8<>
\8</xyps {dup vxp mul exch vyp mul} def % necessita di s>
\8</xyms {dup vxm mul exch vym mul} def % necessita di s>
\8<>
\8</instab {vxp mul dup vyp mul exch vxp mul moveto L0 xyps lineto stroke} def>
\8<>
\8</mmod {div dup truncate sub} def %necessita di a b>
\8<>
\8</punta0 {newpath >
\8<0 0 moveto -10 5 lineto -10 -5 lineto closepath} def>
\8</dirpunta {gsave 2 copy translate 3 -1 roll sub 3 1 roll >
\8<exch sub atan rotate punta0 fill grestore} def>
\8<>
\8<gsave>
\8<150 0 translate>
\8< 0 0 moveto L1 xyps lineto stroke>
\8< 0 0 L1 xyps dirpunta>
\8< 0 0 moveto L1 xyms lineto stroke>
\8< 0 0 L1 xyms dirpunta>
\8< L0 xyms 2 copy moveto L0 xyps exch 3 1 roll add 3 1 roll add exch>
\8< lineto stroke>
\8< L0 xyps 2 copy moveto L0 xyms exch 3 1 roll add 3 1 roll add exch>
\8< lineto stroke>
\8</xydzdw { 4 2 roll 2 copy moveto exch add 3 1 roll add>
\8< exch lineto stroke} def>
\8</instab {vxp mul dup vyp mul neg exch vxp mul 2 copy >
\8<moveto L0 xyps exch 3 1 roll add 3 1 roll add exch lineto stroke} def>
\8</somvett {exch 4 -1 roll add 3 1 roll add} def>
\8<>
\8<4 L mul vyp mul L mmod L mul dup h lt {instab L} if pop>
\8<12 L mul vyp mul L mmod L mul dup h lt {instab L} if pop>
\8<>
\8</A {26} def>
\8</B {80} def>
\8</P1 {A xyms A xyps somvett} def>
\8</P2 {B xyms A xyps somvett} def>
\8</Q1 {A xyms B xyps somvett} def>
\8</Q2 {B xyms B xyps somvett} def>
\8</Z1 {A xyps} def>
\8</Z2 {L0 xyms A xyps somvett} def>
\8</W1 {B xyps} def>
\8</W2 {L0 xyms B xyps somvett} def>
\8<gsave>
\8<2 setlinewidth P1 moveto Q1 lineto stroke>
\8<P2 moveto Q2 lineto stroke>
\8<1 setlinewidth>
\8<[3] 0 setdash Z1 moveto Z2 lineto stroke >
\8<[3] 0 setdash W1 moveto W2 lineto stroke>
\8<%/Helvetica findfont 10 scalefont Z2 moveto (dx) show>
\8<grestore>
\figfin
\*
\eqfig{260pt}{140pt}{%
\ins{170pt}{60pt}{$d\x$}
\ins{150pt}{90pt}{$d\x'$}
\ins{265pt}{62pt}{$\x$}
\ins{235pt}{77pt}{$E$}
\ins{120pt}{22pt}{$\h$}
\ins{95pt}{70pt}{$\h'$}
\ins{50pt}{93pt}{$\h$}}{c2gcpaper}{}
\*
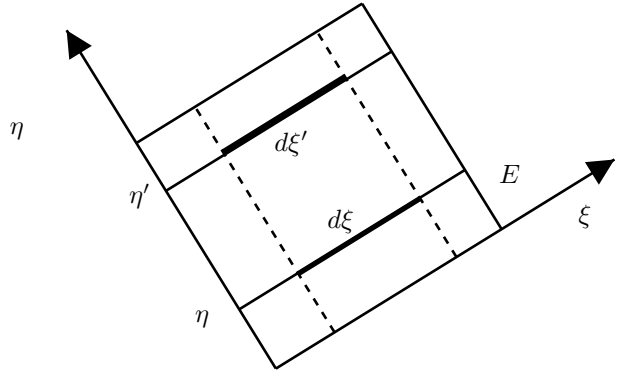
\0{\nota\it Fig. 2: The two lines at constant $\h$ are at height $\h$ and
$\h'$ respectively. The two infinitesimal segments $d\x$ and $d\x'$
correspond to each other as they are crossed by the same set of stable
manifolds, the extreme two of which are drawn as dashed lines.\vfill}
\*

But $\fra{|S^M d\x|}{|S^M d\x'|}\tende{M\to\io}1$ because the two segments
approach each other, while greatly and chaotically erring towards $\io$
on $W^u_O$ at the exponential speed of the expansion rates.

Hence by combining \equ(5.2) for $M\to\io$ and \equ(4.3), we see that
the ratio of the measures $\lis \r^\h(\x)d\x/\lis \r^{\h'}(\x')d\x`$ in
corresponding intervals $d\x,d\x'$ near the corresponding points
$x,x'\in E$ with local coordinates $(\x,\h)$ and $(\x,\h')$, is simply:
$$\prod_{-\io}^{+\io}\fra{\L_u^{-1}(S^{-j}x)}{\L_u^{-1}(S^{-j}x')}
\Eq(5.3)$$
This shows that the SRB distribution $\m$ can be formally given by
attributing to the ``points'' on the unstable manifold of $A$ a weight
given by:
$$const\,\prod_{-\io}^{+\io}\L_u^{-1}(S^{-j}x)=const\,
e^{-\sum_{-\io}^{+\io}\log \L_u(S^{-j}x)}\Eq(5.4)$$
or by a density on $W^u_O$ given formally by the product in \equ(4.3).
Such statements should be interpreted in the same way in which one
interprets statements like: "the one dimensional Ising model with
nearest neighbour interaction attributes to the spin configuration
$(\s_i)_{-\io}^\io$ the probability":
$$const\,e^{-\sum_{-\io}^\io J\s_i\s_{i+1}}\Eq(5.5)$$
The formal expression \equ(5.4) must, therefore, be interpreted as a
limiting statement.  The function $h(x)\=\log\L_u(x)$ in \equ(5.4) plays
the same role as $J \s_0\s_{1}$ in the Ising model in \equ(5.5) and the
appropriate way of understanding \equ(5.4) is, as we have in fact
discussed, as a limit of \equ(4.3).  The important realization of Sinai,
[S2], was that the equation \equ(5.4), via \equ(5.5), had a close analogy
in statistical mechanics.\footnote{${}^{12}$}{\nota Hence the name of
"thermodynamic formalism" given by Ruelle to the mathematical theory
based on the above point of view, [R1].}

This remark led Sinai to his general theory of Markov partitions (see
below) which are the main technical tool that is used to show
mathematically that the limit of \equ(4.2) or (more precisely) the limit
in \equ(4.5) really exist and for describing its general properties in
satisfactory detail.  \*

\0{\it\S6 Coarse graining and Markov partitions.}
\numsec=6\numfor=1 \*

The above discussion has, as already said at the beginning of \S5, just
heuristic value as it has not led to a really usable formula for
$\lis\r^\h(\x)$, but just to a few relations that such function must
obey when evaluated on $W^u_O$.

The solution lies in a stricter interpretation of the thermodynamic
analogy (\ie the similarity between \equ(5.4) and \equ(5.5)).  To
understand the rigorous solution (given in [S2]) to the problem of
showing the existence of the limit \equ(4.5) and the existence of
$\lis\r^\h(\x)$ and $\n(d\h)$, one has to introduce the concept of a
``parallelogram'' and of a {\it Markov partition $\EE$} of the phase
space $\CC$ into parallelograms. This can be ultimately related to the
problem of constructing a good division of the phase space in cells (\ie
a ``good'' coarse graining) so that the evolution can be correctly
represented as a cell permutation, without ``distorting'' the hyperbolic
nature of the motion (for such an interpretation of what follows see
[Ga4]).

A parallelogram will be a small set with a boundary consisting of pieces
of the stable and unstable manifolds joined together as described
below. The smallness has to be such that the parts of the manifolds
involved look essentially ``straight'': \ie the sizes of the sides have
to be small compared to the smallest radii of curvature of the manifolds
$W^u_x$ and $W^s_x$, as $x$ varies in $\CC$.

Therefore let $\d$ be a length scale small compared to the minimal
(among all $x$)
curvature radii of the stable and unstable manifolds. Let
$W^{u,\d}_x,W^{s,\d}_x$ be the connected parts of $W_x^u$, $W^s_x$
containing $x$ and contained in a sphere of radius $\d$.

Let us first define a {\it parallelogram} $E$ in the phase space $\CC$,
to be denoted by $\D^u\times\D^s$, with center $x$ and axes $\D^u$,
$\D^s$ with $\D^u$ and $\D^s$ small connected surface elements on
$W^u_x$ and $W^s_x$ containing $x$. Then $E$ is defined as
follows. Consider $\x\in\D^u$ and $\h\in\D^s$ and suppose that the
intersection $\x\times\h\=W^{s,\d}_\x\cap W^{u,\d}_\h$ is a unique point
(this will be so if $\d$ is small enough and if $\D^u$, $\D^s$ are small
enough compared to $\d$ as we can assume, because the stable and
unstable manifolds are ``smooth''\footnote{${}^{13}$}{\nota This is only
approximately true because they are H\"older continuous with some
positive exponent, related to the gap between the positive or negative
Lyapunov exponents and $0$.} and transversal, see footnote ${}^1$).

The set $E=\D^u\times\D^s$ of all the points generated in this way when
$\x,\h$ vary arbitrarily in $\D^u,\D^s$ will be called a parallelogram
(or rectangle), if the boundaries $\dpr\D^u,\dpr\D^s$ of $\D^u$and
$\D^s$ as subsets of $W^u_x$ and $W^s_x$, respectively, have zero
surface area on the manifolds on which they lie.  The sets $\dpr_u
E\=\D^u\times\dpr\D^s$ and $\dpr_s E=\dpr\D^u\times\D^s$ will be called
the {\it unstable} or {\it horizontal} and {\it stable} or {\it
vertical} sides of the parallelogram $E$.

\figini{c3gcpaper}
\8</punto { gsave >
\8<3 0 360 newpath arc fill stroke grestore} def>
\8</puntino { gsave >
\8<2 0 360 newpath arc fill stroke grestore} def>
\8</origine1assexper2pilacon|P_2-P_1| { >
\8<4 2 roll 2 copy translate exch 4 1 roll sub >
\8<3 1 roll exch sub 2 copy atan rotate 2 copy >
\8<exch 4 1 roll mul 3 1 roll mul add sqrt } def>
\8<>
\8</punta0{0 0 moveto dup dup 0 exch 2 div lineto 0 >
\8<lineto 0 exch 2 div neg lineto 0 0 lineto fill >
\8<stroke } def>
\8<>
\8</dirpunta{>
\8<gsave origine1assexper2pilacon|P_2-P_1| >
\8< 0 translate 7 punta0 grestore} def>
\8<>
\8</uno{% a v>
\8<dup mul div neg} def>
\8</due{% T v b>
\8<sub div} def>
\8</p{% a b T v>
\8<dup 5 1 roll 3 -1 roll due 3 1 roll exch uno add} def>
\8<gsave>
\8<40 40 40 0 360 arc>
\8<20 40 moveto 60 40 lineto>
\8<40 20 moveto 40 60 lineto>
\8<stroke>
\8<140 40 40 0 360 arc >
\8<120 40 moveto 160 40 lineto>
\8<140 20 moveto 140 60 lineto>
\8<stroke>
\8<150 40 2 0 360 arc fill>
\8<140 50 2 0 360 arc fill>
\8<150 50 2 0 360 arc fill>
\8<stroke>
\8<175 75 moveto 160 60 lineto>
\8<175 75 160 60 dirpunta>
\8<150 10 moveto 150 70 lineto>
\8<110 50 moveto 165 50 lineto>
\8<stroke>
\8<>
\8<240 40 40 0 360 arc stroke>
\8<[3] 0 setdash 220 20 moveto 220 60 lineto stroke>
\8<220 60 moveto [1] 0 setdash 260 60 lineto stroke>
\8<[3] 0 setdash 260 60 moveto 260 20 lineto stroke>
\8<[1] 0 setdash 260 20 moveto 220 20 lineto stroke>
\8<220 40 moveto 260 40 lineto 240 20 moveto 240 60 lineto>
\8<stroke>
\8<220 40 puntino>
\8<260 40 puntino>
\8<240 60 puntino>
\8<240 20 puntino>
\8<grestore>
\figfin

\eqfig{260pt}{90pt}{
\ins{43pt}{37pt}{$x$}
\ins{43pt}{60pt}{$\D^s$}
\ins{60pt}{40pt}{$\D^u$}
\ins{155pt}{36pt}{$\x$}
\ins{130pt}{60pt}{$\h$}
\ins{177pt}{77pt}{$\x\times\h$}
\ins{245pt}{70pt}{$E$}
}{c3gcpaper}{}
\*
{\nota
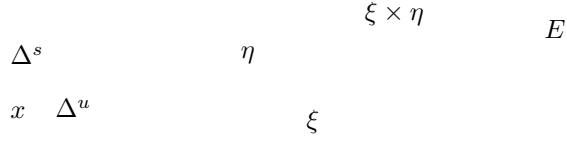
\0Fig. 3: The circles are a neighborhood of $x$ of size very small
compared to the curvature of the manifolds; the first picture shows the
axes; the intermediate picture shows the $\times$ operation and
$W^{u,\d}_\h, W^{s,\d}_\x$ (the segments through $\h$ and $\x$ have size
$\d$); the third picture shows the rectangle $E$ with the axes and the
four marked points are the boundaries $\dpr\D^u$ and $\dpr\D^s$. The
picture refers to a two dimensional case and the stable and unstable
manifolds are drawn as flat, \ie the $\D$'s is very small compared to
the curvature of the manifolds.\vfill}
\*
Consider now a partition $\EE=(E_1,\ldots,E_\NN)$ of $\CC$ into $\NN$
rectangles $E_j$ with pairwise disjoint interiors. We call
$\dpr_u\EE\=\cup_j\dpr_u E_j$ and $\dpr_s\EE\=\cup_j\dpr_s E_j$: these
are called respectively the {\it unstable boundary} of $\EE$ and the
{\it stable boundary} of $\EE$, or also the horizontal and vertical
boundaries of $\EE$, respectively.

We say that $\EE$ is a {\it Markov partition} if the transformation $S$
acting on the stable boundary of $\EE$ maps it into itself (in formula
this is: $S \dpr_s\EE\subset \dpr_s\EE$) and if likewise the map
$S^{-1}$ acting on the unstable boundary maps it into itself
($S^{-1}\dpr_u \EE\subset \dpr_u\EE$).

%Attempting to draw what this means greatly clarifies the matter.

The actual construction of the SRB distribution then proceeds from the
important result of the theory of Anosov systems expressed by
what we shall call ``Sinai's first theorem'': \*

{\it Theorem: every transitive Anosov system admits a Markov partition}
$\EE$, [S2].

\* The above theorem is the first step towards a controlled version of
the heuristic arguments given above and towards a usable form of
equation \equ(4.5) based on a suitable interpretation of \equ(5.3).  It
can be extended to imply the existence of more special Markov
partitions: for instance to show the existence of Markov partitions with
any one of the following three properties (the last shows that the first
two can be realized simultaneouly and will play a key role in our
analysis):

1) The construction of $\EE$ can be done, [Ga3], so that the horizontal
axes of $E_j$ {\it all lie on} $W^u_O$ (and the vertical on $W^s_O$) and
their union is a set that can be obtained from a single small connected
surface element $\lis\D$ of $W^u_O$ (resp. $\lis\D'$ of $W^s_O$)
containing $O$ by dilating it with a high iterate $S^Q$ of the time
evolution $S$. In other words the union $\cup_j\D^u_j$ of the horizontal
axes of the parallelograms $E_j\in\EE$ can be regarded as a good finite
approximation to our attractor $A$, because it has the form $
S^Q\lis\D$ with $\lis\D$ a connected surface element of the unstable
manifold $W^u_O$, containing $O$. Likewise the union of the stable
axes can be regarded as a large connected part of the stable manifold
$W^s_O$.

2) If the reversibility property holds it is clear that $i\EE$ is also a
a Markov partition. This follows from the definition of Markov
partition and from the fact that reversibility implies:

$$W^s_{x}=i W^u_{ix}\Eq(6.1)$$

\0The definition of a Markov partition also implies that the intersection
of two Markov partitions is a Markov partition, hence it is clear that
there are Markov partitions $\EE$ that are reversible in the sense that
$\EE=i\EE$.

3) Furthermore one can construct a Markov partition $\EE$ which
is reversible and at the same time verifies the property 1) above, [Ga3].

Here we shall use Markov partitions that verify property 3) above.
\*

In order to formulate Sinai's second theorem, which gives an expression
for a controlled approximation to the SRB distribution, we consider the
partition $\EE_T=\cap_{-T}^T S^{-j}\EE$ obtained by intersecting the
images under $S^j$, $j=-T,\ldots,T$ of $\EE$. Then $\EE_T$ is still a
Markov partition and it is time reversal invariant if $\EE$ is.  We now
construct a probability distribution that we regard concentrated on the
finite approximation $A_{T}$ to the attractor, consisting of the union
of the horizontal axes of $\EE_T$, (see remark 1),2) above), and equal,
with the notations of remark 1 above, to $S^{T+Q}\lis\D=A_{T}$.

We can visualize the small parallelograms forming $\EE_T$ as a lattice
of parallelograms: two parallelograms adjacent and on the same vertical
strip will have horizontal axes that correspond to each other in the
same sense that the close horizontal surface elements used in deriving
\equ(5.3) correspond. Therefore if we attribute to the horizontal axis
of the parallelogram $E_j$ in $\EE_T$ with center $x_j$ a weight equal
to $\prod_{h=-\t/2}^{\t/2} \L_u^{-1}(S^h x_j)$ {\it we see that the
ratios of the weights of corresponding surface elements automatically
realize an approximation of the product in \equ(5.3)}. We take, of
course, $\t\ll T$ so that the size of each parallelogram is so small
that the weight we attribute to each does not depend on which point of
$E_j$ we regard as a center and that no essential ambiguity arises as to
which weight to attibute to a parallelogram.\footnote{${}^{14}$}{\nota
The size of the parallelograms of $\EE_T$ is clearly decreasing as
$e^{-\l n}$, at least, if $\l$ is the spectral gap, see footnote ${}^1 $
above.} Note that the above weight $\lis\L^{-1}_{u,\t}(x)$ is the
inverse of the expansion coefficient of the map $S^\t$ as a map of
$W^u_{S^{-\t/2}x}$ to $W^u_{S^{\t/2}x}$ (between $S^{-\t/2}x$
and $S^{\t/2 }x$), \ie:
$$\lis\L_{u,\t}(x)=\prod_{j=-\t/2}^{\t/2-1}\L_u(S^jx)\Eq(6.2)$$
A similar quantity can be defined by regarding $S^\t$ as a map of
$W^s_{S^{-\t/2}x}$ to $W^s_{S^{\t/2}x}$.

The construction thus generates a probability distribution which, by the
above analysis, verifies \equ(5.3) more and more exactly as $\t\to\io$.
Hence this analysis suggests the following theorem (which we shall call
``Sinai's second theorem'', [S2]):

\* \0{\it Theorem: If $(\CC,S)$ is a transitive Anosov system the SRB
distribution $\m$ exists and the $\m$ average of a smooth function
$F$ is:
$$\ig_\CC \m(dx) F(x)=\lim_{T\to\io,\t\to\io} \fra{\sum_j
\lis\L^{-1}_{u,\t}(x_j) F(x_j)}{\sum_j
\lis\L^{-1}_{u,\t}(x_j)}{\buildrel def \over
=}\lim_{T\to\io,\t\to\io}\ig_\CC \m_{T,\t}(dx) F(x) \Eq(6.3)$$

\0where, with the above notations, $x_j$ is a point in $E_j\in \EE_T$.
}\*

The above $\m_{T,\t}$ as defined by the middle ratio in \equ(6.3),
can be taken as a ``concrete'' procedure to follow in approximating the
SRB distribution.

In the case of equilibrium under assumption (C) in \S2 the distribution
$\m$ in equation \equ(6.3) can be shown to coincide with the
microcanonical ensemble, as already mentioned,
[S1].\footnote{${}^{15}$}{\nota One should not be disturbed by the fact
that this is a rigorous mathematical theorem only for Anosov systems or,
more generally, for ``axiom A'' attractors, [R1]: one should not forget
that the microcanonical ensemble is also lacking a mathematical
justification in equilibrium theory. In fact the only equilibrium case
in which one can prove the ergodic hypothesis is for the Lorentz gas
(\ie the {\it billiards}), [S1], [BSC], with $N=1$: in such a case
($N=1,\,\f=0$ and $\f^u$ a suitable hard core potential, \ie a
triangular lattice of hard disks) the present point of view can also be
shown to hold in the presence of dissipation, [CELS].}

\vskip1.cm
\0{\it\S7 Application.}
\*
\numsec=7\numfor=1

The chaotic hypothesis can be taken as an extension of the ergodic
hypothesis for equilibrium statistical mechanics to systems in
nonequilibrium stationary states (conservative or dissipative). In the
equilibrium (\ie conservative) case it implies the ergodic hypothesis
({\it but it is stronger}) and hence the microcanonical distribution,
which we know how to use in order to draw physical consequences.

It is therefore legitimate to ask whether the chaotic hypothesis and the
ensuing SRB distribution have any predictive value of their own. Just as
the ergodic hypothesis implies the well tested classical thermodynamics,
the new hypothesis should imply, for example, irreversible thermodynamics
of nonequilibrium stationary states, without the necessity of solving
the equations of motion.  It is not clear that this is so.

However there are already some experimental results that offer support
to the chaotic hypothesis, since one can understand their outcome by
using it.

Here we examine, in particular, one experimental result, [ECM2], which
the authors already attempted to explain by relating it to our chaotic
hypothesis.  Some of the data in [ECM2] require, to be unambigously
understood, the discussion in [ES].  We shall take the viewpoint of the
preceding sections to make more precise, and detailed, the argument in
[ECM2], modifying it to some extent in order to put it on a more
mathematical basis.

In the context of this paper the experiment [ECM2] deals with model 2 in
\S3 and measures the entropy production rate (\ie phase space
contraction rate, see \S3, \equ(3.5)) as seen on a stretch of time $\t$,
short compared to the duration of the experiment $T$, and repeating the
measurement $T/\t$ times.\footnote{${}^{16}$}{\nota The reader should
not mind that the symbol for the integer $T$ is sometimes also used for
the absolute temperature.} We emphasize that this is an
experiment on a system {\it far} from equilibrium.

Calling $D\s_\t(x)$ the entropy production rate measured on the motion
originating at $S^{-\t/2}x$ and observed $\t$ units of time (we take
$\t$ even for simplicity), we define it, see \equ(3.4),\equ(3.5), by:
$$D\s_\t(x)=D\,\fra1{\t}\sum_{j=-\t/2}^{\t/2-1}{\s(S^{j}x)}\
{\buildrel def \over =}\ D\,\media{\s}_+\, a_\t(x)\Eq(7.1)$$

\0where $\media{\s}_+$ is the average in the future of $\s(S^jx)$, which
is a constant almost everywhere in phase space with respect to
$\m_0$--random choices of initial data.\footnote{${}^{17}$}{\nota This
because the SRB distribution verifies the extended zeroth law, \equ(2.1),
which says that the averages are, with $\m_0$--probability $1$,
independent of the initial data.}

The total entropy production while the phase space point $x$ evolves
between $S^{-\t/2}x$ and $S^{\t/2}x$ is obtained by multiplying
\equ(7.1) with the time elapsed during $\t$ such timing collisions.  For
simplicity we think that the {\it time interval $t_0$} between the
timing collisions is constant. Note that $x$ is the middle point of the
segment of a trajectory of (discrete) time length $\t$, defining the
fluctuation $a_\t(x)$ in \equ(7.1).

{\it It is perhaps important to stress that $\media{\s}_+$ is very
different from the limit as $\t\to+\io$ of $\s_\t(x)$ in \equ(7.1): in
fact the latter (by the time reversal symmetry) vanishes, while the
former is positive as follows from numerical evidence, see \S3, and
as assumed in (A) in \S2.}

The experiment divided the $a_\t$--axis into small intervals
$I_0,I_{\pm1},\ldots$ and measured the quantity $a_\t(x)$ building a
histogram counting how many times the $a_\t$ fell into the interval
$I_p$ (where $a_\t(x)=p$).  Obviously we expect a distribution
$\p_\t(p)$centered around an average which is, (see \equ(3.4) and
\equ(7.1)), exactly $1$.  The result for $\p_\t(p)$ can be found in fig. 1
of [ECM2], for one (rather large) value of $\t$ and one of $\g$ and
$N=56$.

A second experimental result is for $\P_\t(p)=-\fra1{2
N\t t_0\media{\s}_+}\log\fra{\p_\t(p)}{\p_\t(-p)}$, \ie essentially for the
logarithm of the ratio of the probability that $a_\t(x)=p$ to that of
$a_\t(x)=-p$. The result, fig.  2 of [ECM2] is, for the rather large
value of $\t$ considered, a remarkably precise straight line for
$\P_\t(p)$ as a function of $p$, \ie $\P_\t(p)$ is a linear function of
$p$.

A third experiment shows that the slope of this line as a function of
$\t$ verifies, even for large deviations, the relation
of proportionality to $\t$ for $\t$ large (fig. 3 of [ECM2]).

The results are rather precise with apparently little margin for errors,
hence one has to find a theoretical reason that the
probability distribution of $D\,\s_\t(x)$ has the form:
$$\eqalign{
\p_\t(p)\,dp{\buildrel def \over =}&
P(a_\t\in(p,p+dp))=e^{-\t\z(p)+\t Cp}dp\qquad\quad {\rm or} \cr
\fra{\p_\t(p)}{\p_\t(-p)}=& e^{2\t C p}\cr}\Eq(7.2)$$
for a suitably chosen constant $C$ and a suitable even function $\z(p)$
with minimum at $p=1$ and with the argument of the exponential correct
up to, apparently, $p,\t$ independent corrections (see fig. 3 of
[ECM2]). In [ECM2] a theoretical argument is presented which leads
to $2C= D t_0 \media{\s}_+ $, if $t_0$ is the average time between
timing events.

We are now going to show, and this is our main technical result (and a
theorem under assumptions (A,B,C) of \S2), what we call a {\it
fluctuation theorem}: \*

\0{\it Fluctuation theorem: Let $(\CC,S)$ verify the properties (A,B,C)
of \S2, (dissipativity, reversibility and chaoticity). Then the
probability $\p_\t(p)$ that the total entropy production $D\t t_0\s_\t(x)$,
\equ(7.1), over a time interval $t=\t t_0$ (with $t_0$ equal to the
average time between timing events) has a value $Dt\media{\s}_+ p$
verifies the large deviation relation:

$$\fra{\p_\t(p)}{\p_\t(-p)}= e^{Dt \media{\s}_+ p}\Eq(7.3)$$

\0with an error in the argument of the exponential which can be
estimated to be $p,\t$--in\-depen\-dent.}\*

This means that if one plots the logarithm of the left hand side of
\equ(7.3) as a function of $p$ one observes a straight line with more
and more precision as $\t$ becomes large (in agreement with figure 3 in
[ECM2]).  \*

{\it Remark: since the above theorem is deduced under the assumptions
(A,B,C) only, the result \equ(7.2) will apply as well to the models
1,3,4,5. This gives a parameterless prediction of the outcome of several
numerical experiments similar to the one described above.}  \*

The main ideas for the proof, [CG], [Ga3], [Ga2], of the above theorem
are the following.

The probability that $a_\t(x)\in I_p$ over the probability that
$a_\t(x)\in I_{-p}$ is, if one uses the notations and the approximation
$\m_{T,\t}$ to $\m$ described at the end of \S4 (see \equ(6.3)) with
$F(x)=a_\t(x)$:
$$\fra{\sum_{j,\,a_\t(x_j)=p} \lis\L_{u,\t}^{-1}(x_j)}
{\sum_{j,\, a_\t(x_j)=-p} \lis\L_{u,\t}^{-1}(x_j)}\Eq(7.4)$$
where $\lis\L_{u,\t}(x)$ is the jacobian determinant of $S^\t$ as a map
of $W^u_O$ into itself, evaluated at the point $S^{-\t/2}x\in S^T\D$,
(\ie as a map between $S^{-\t/2}x$ and $S^{\t/2}x$, see \equ(6.2),
\equ(7.1)).

Since $\m_{T,\t}$ in \equ(6.3) is only an approximation at fixed $T,\t$
{\it an error is involved in using \equ(7.4)}. It can be shown that this
error can be estimated to affect the result only by a factor bounded
above and below uniformly in $\t,p$, [CG], [Ga3], [Ga2]. This is a
remark technically based on the thermodynamic analogy pointed out in
\equ(5.4), \equ(5.5).

We now try to establish a one to one correspondence between the addends
in the numerator of \equ(7.4) and the ones in the denominator,
aiming at showing that corresponding addends have a {\it constant
ratio} which will, therefore, be the value of the ratio in \equ(7.4).

This is possible because of the reversibility property (B), \S2.  Let
$x\in A$, then $ix\in iA$.  By using the identity $S^{-\t}(S^\t
x)=x$, the identity $S^{-\t}(i S^{-\t}x)=ix$ (time reversal) and
\equ(6.1), we deduce the relations:\footnote{${}^{18}$}{\nota The key
remark is that time reversal $i$ maps $E_j$ into $iE_j$ and at the
same time {\it changes} the horizontal surface elements of $E_j$ into
the vertical ones of $iE_j$ and the vertical surface elements of $E_j$
into the horizontal of $iE_j$, see \equ(6.1). }
$$a_\t(x)=-a_\t(ix),\qquad
\lis\L_{u,\t}(ix)=\lis\L_{c,\t}^{-1}(x)
\Eq(7.5)$$
which are identities, see [ECM2],[CG],[Ga3]. The first equality in
\equ(7.5) is obvious as in all the cases considered the $i$ operation
changes the sign to $\s(x)$, the rate of change of the phase space
volume. The second equality in \equ(7.5) is also easy to check: in fact
let $\b$ be a surface element on $W^u_x$ around $S^{-\t/2}x$ and let
$\b'=S^\t\b$ be its $S^\t$ image around $S^{\t/2}x$: then
$\lis\L_{u,\t})x)=\fra{|\b'|}{|\b|}$. Applying $i$ to $\b$ and $\b'$ one
obtains surface elements $i\b$ and $i\b'$ on $W^s_{ix}$ with the same
area as $\b$ and $\b'$ (because $i$ is an isometry) around respectively
$S^{\t/2}ix$ and $S^{-\t/2}ix$ so that the expansion rate
$\lis\L_{s,\t}(ix)$ is $\fra{|i\b|}{|i\b'|}
=\fra{|\b|}{|\b'|}=\lis\L^{-1}_{u,\t}(x)$.

The ratio \equ(7.4) can therefore be written simply as:
$$\fra{\sum_{E_j, a_\t(x_j)=p} \lis\L_{u,\t}^{-1}(x_j)}
{\sum_{E_j, a_\t(x_j)=-p} \lis\L_{u,\t}^{-1}(x_j)}
\=\fra{\sum_{E_j, a_\t(x_j)=p} \lis\L_{u,\t}^{-1}(x_j) }
{\sum_{E_j, a_\t(x_j)=p} \lis\L_{c,\t}(i x_j)}\Eq(7.6)$$
where $x_j\in E_j$ is a point in $E_j$. In deducing the second relation
we make us of the existence of the time reversal symmetry $i$ and of
\equ(7.5).

It follows then that the ratios between corresponding terms in the ratio
\equ(7.6) is equal to $\lis\L_{u,\t}^{-1}(x)\lis\L_{c,\t}^{-1}(x)$.
This differs from the reciprocal of the total variation of phase space
volume over the time $\t$ between the point $S^{-\t/2}x$ and $S^{\t/2
}x$ only because it does not take into account the ratio of the sines of
the angles $\th(S^{-\t/2}x)$ and $\th(S^{\t/2}x)$ formed by the stable
and unstable manifolds at the points $S^{-\t/2}x$ and $S^{\t/2}x$, see
footnote ${}^1$.  But $\lis\L_{u,\t}^{-1}(x)\lis\L_{c,\t}^{-1}(x)$ will
differ from the actual phase space contraction under the action of
$S^\t$, as a map between $S^{-\t/2}x$ and $S^{\t/2}x$, by a factor that
can be bounded between $B^{-1}$ and $B$ with
$B=\max_{x,x'}\fra{|\sin\th(x)|}{|\sin\th(x`)|}$ which is finite by the
transversality of the stable and unstable manifolds.

Now for all the points $x_j$ in \equ(7.6), the reciprocal of the total
phase space volume contraction over a time $t_0$ is
$e^{-a_\t(x_j)\media{\s}_+t_0\t D}$, which (by the constraint imposed on
the summation labels $a_\t=p$) equals
$e^{-Dt_0\t\media{\s}_+\,p}$. Hence the ratio \equ(7.4) will be
$e^{Dt_0\t\media{\s}_+\,p}$, to leading order as $N,\t\to\io$, proving
\equ(7.3), with $2C=D\media{\s}_+t_0$. It is important to note that
there are two errors ignored here, as pointed out in the discussion
above.\annota{{19}}{In the previous paragraph and in the paragraph following
\equ(7.4).} They imply that the argument of the exponential {\it is
correct up to $p,\t$ independent corrections} (which are in fact
observed in the experiment as fig.3 of [ECM2] shows).

The $p$ independence of the coefficient of $C$ in \equ(7.2) is therefore
a key test of the theory (and it should hold with corrections of order
$O(\t^{-1})$).

\vskip1cm
\0{\it\S8 Outlook.}
\* \numsec=8\numfor=1

We end with a number of remarks.
\*

\01) The interest of our discussion in \S7 is not, of course, the
fluctuation theorem which is essentially proved there (for a formal
proof see [Ga2] and [Ga3]), but in the clarification of the meaning of
the properties (A,B,C) mentioned in \S2. Furthermore it is interesting
that our chaotic hypothesis, \S2 does have some concrete and
experimentally verifiable consequences (verified here in the case of model
2): such consequences have the remarkable feature of being predictions
{\it without free parameters}, hinting that the hypothesis might have a
quite general validity.  One cannot be too demanding on the matter of
mathematical rigor: one should not forget that the ergodic hypothesis is
far from being proved either, particularly in the generality one would
want.

\* \02) The fluctuation formula \equ(7.3) holds also for the models
1,3,4,5 because the fluctuation theorem applies to all such models (see
remark following the theorem): but the numerical experiments do not seem
to exist yet.  \*

\03) The pairing property (D) and the smooth distribution of the
Lyapunov exponents (E) have been used here only to get some intuition
and to visualize the hyperbolic nature of the attractor and the equality
of the dimensions of the stable and unstable manifolds.  It seems
interesting to perform numerical experiments to try to investigate
better, at least in the systems that we are considering, if the density
function $f_\io(x)$ is really positive at $x=0$, see (E), \S2, as the
numerical results seem to suggest in some cases, [LPR], [ECM1], [SEM].
\*

\04) Note that the fluctuation theorem (\equ(7.3)) applied to model 5
leads to an interesting consequence on the large deviations properties
of the magnitude of the energy dissipation $\e$ in turbulent flows. In
this case we take, for simplicity, the kinetic energy $D_v k T(x)$ of
the viscous modes to be a non fluctuating quantity equal to $D_vkT$. Then
the random variable $p$ associated with $\s(x)=\fra{\e(x)}{D_v k T(x)}$
in the fluctuation theorem of \S7 is just proportional to the average
over a time interval $t$ of the energy dissipation rate $\e$. This is a
variable that is assumed to be constant in the Kolmogorov--Obuchov
theory: what we say here is that it is in fact a fluctuating quantity
and we predict (on the basis of the fluctuation theorem of \S7)
that the time average $\media{\e}_t$,
over a time interval $t$, of $\e(x)$,\ie ${\media{\e}}_t\= t D_v
\media{\e}_+ p$ is such that its probability distribution $\p_t(p)$
verifies the {\it linear large deviation law}:

$$\fra1{D t p\media{\e}_+/kT}\log
\big[\p_t(p)/\p_t(-p)\big]\=1\Eq(8.1)$$

\0up to corrections of order $O(t^{-1})$. If $T=T(x)$ has to be regarded
as a fluctuating variable then \equ(8.1) must be regarded as a property
of the fluctuations of the entropy production rate $\s(x)=\fra{\e(x)}{k
T(x)}$ rather than of the energy dissipation, (with some obvious
modifications, \eg $\media{\e}_+/kT\to\media{\e/kT}_+$).  \*

\05) Concerning the particularity of the gaussian thermostat we think,
see [Ga1],[C], that there should be, also in nonequilibrium, several
equivalent ways of describing the same stationary distribution
corresponding to different $ \m$ and to different physical ways of
reaching the stationary state, at least in the thermodynamic limit. And
it may well be that the gaussian thermostat turns out to be equivalent
to other models of thermostats, which could be described by rather
different attractors.  For instance a stochastic
thermostat, in which a particle colliding with the wall is scattered
with a maxwellian distribution at a given temperature, will certainly be
described by a statistics $\m$ which is absolutely continuous with
respect to the Liouville distribution.\footnote{${}^{20}$}{\nota Note that a
stochastic model of thermostat is described by a stochastic differential
equation and therefore our discussion does not apply without some major
modification.}  In the thermodynamic limit this might just give the same
result as obtained with a statistics which, for finite $N$, is on a
fractal attractor.  This mechanism is like the one realized by the
microcanonical and the canonical ensembles (the first is concentrated on
a set of configurations which has zero probability with respect to the
second, as long as $N<\io$).  This is clearly a question that requires
further investigations.  \*

\06) One can also regard the gaussian thermostat as a device to
eliminate some trivial Lyapunov exponents. For instance in model 4 we
could simply {\it not} introduce the gaussian constraint that the {\it
total} energy is constant (which led to the $\a_0\V p_j$ terms): we
believe that physically the system would then still behave in the same
way, {\it for large $N,L$}. But without such a constraint we could not
assume the phase space to have $4N-3$ dimensions, because $H$ would not
be rigorously constant. We expect, however, that in such case the energy
$H$ is {\it approximately} constant, in fact more and more so as
$L,N\to\io$ with $NL^{-2}=n$ constant. Thus the variability of $H$
probably leads to a zero Lyapunov exponent making the chaoticity
assumption manifestly invalid. But this would be so only in a somewhat
trivial way as its violation is due to a zero Lyapunov exponent
associated with a variable which is ``almost'' a constant of the motion.
Hence it is natural to fix the value of the energy rigorously \ap by a
constraint (realized via a minimum constraint principle, like Gauss'
principle) and so dispose of the extra $0$ (or very close to $0$)
Lyapunov exponent, recovering then again a situation in which the system
is strictly chaotic. Such a point of view can be extended to cover cases
in which hyperbolicity is not valid because of the existence of quasi
exact conservation rules. An example is in fact model 2 in which the
variables $\tilde y_j,\tilde p_{yj}$ can be replaced by $y_j, p_{yj}$
thus turning $P_{x},X_{x}$ into {\it variable} quantities: their
variability is however clearly due to the special (vertical) boundary
conditons used and it should therefore not matter whether they are kept
rigorously constant or not, in the limit of $N,L\to\io$.  The dynamics
can be modified by turning such quantities into exact conservation laws
and the new dynamics should be indistinguishable from the previous one
in the thermodynamic limit. Another example is provided by the
constraints imposed on model 3 to achieve that the horizontal momentum
is conserved.

\07)  Like for the ergodic hypothesis in equlibrium, the range of
validity of the chaotic hypothesis for nonequilibrium stationary states
is not known; the more complicated the nature of the latter states,
maintained in the presence of external fields or special boundary
conditions, makes this case even more difficult.
\*
\0{\it Aknowledgements:} We are indebted to J.L. Lebowitz and
G. Eyink for clarifying and patient explanations on their papers [CELS]
and [CL], even before publication. We are also indebted to Y. Sinai for
stimulating discussions.\\ G.G. is indebted to Rutgers University and to
Rockefeller university for partial support and to CNR-GNFM for travel
support. E.G.D.C. acknowledges financial support under contract
DE-FG02-88-ER13847 of the US Department of Energy.

\vskip1cm
{\bf References.}
\*
\item{[AA]} Arnold, V., Avez, A.: {\it Ergodic problems of classical
mechanics}, Benjamin, 1966.

\item{[Bo]} Bowen, R.: {\it Equilibrium states and the ergodic theory of
Anosov diffeomorphisms}, Lecture notes in mathematics, vol. {\bf 470},
Springer Verlag, 1975.

\item{[BEC]} Baranyai, A., Evans, D.T., Cohen, E.G.D.: {\it
Field dependent conductivity and diffusion in a two dimensional Lorentz
gas}, Journal of Statistical Physics, {\bf70}, 1085-- 1098, 1993.

\item{[BSC]} Bunimovitch, L., Sinai, Y.G., Chernov, N.I.: {\it Statistical
properties of two dimensional hyperbolic billiards}, Russian
Mathematical Surveys, {\bf 45}, n. 3, 105--152, 1990.

\item{[C]} Cohen, E.G.D.: {\it Boltzmann and statistical mechanics},
Lecture delivered at the ``Interbational Conference on Boltzmann and its
legacy 150 years after his birth'', Accademia Nazionale dei Lincei, Roma
1994, May 21--28, in print.

\item{[CELS]} Chernov, N.I., Eyink, G.L., Lebowitz, J.L., Sinai, Y.G.:
{\it Steady state electric conductivity in the periodic Lorentz gas},
Communications in Mathematical Physics, {\bf 154}, 569--601, 1993.

\item{[CG]} Cohen, E.G.D., Gallavotti, G.: {\it Dynamical ensembles in
nonequilibrium statistical mechanics}, archived in $mp\_
arc@math.utexas.edu$, \#94-340; to appear in Physical Review Letters,
1995.

\item{[CL]} Chernov, N.T., Lebowitz, J.L.: {\it Stationary shear flow in
boundary driven hamiltonian systems}, Rutgers University preprint, may
1994.

\item{[Dr]} Dressler, U.: {\it Symmetry property of the Lyapunov
exponents of a class of dissipative dynamical systems with viscous
damping}, Physical Review, {\bf 38A}, 2103--2109, 1988.

\item{[ECM1]} Evans, D.J.,Cohen, E.G.D., Morriss, G.P.: {\it Viscosity of a
simple fluid from its maximal Lyapunov exponents}, Physical Review, {\bf
42A}, 5990--\-5997, 1990.

\item{[ECM2]} Evans, D.J.,Cohen, E.G.D., Morriss, G.P.: {\it Probability
of second law violations in shearing steady flows}, Physical Review
Letters, {\bf 71}, 2401--2404, 1993.

\item{[ER]} Eckmann, J.P., Ruelle, D.: {\it Ergodic theory of strange
attractors}, Reviews of Modern Physics, {\bf 57}, 617--656, 1985.

\item{[ES]} Evans, D.J., Searles, D.J.: {\it Equilibrium microstates which
generate second law violating steady states}, Research school of
chemistry, Canberra, ACT, 0200, preprint, 1993.

\item{[Ga1]} Gallavotti, G.: {\it Ergodicity, ensembles, irreversibility
in Boltzmann and beyond}, preprint Physics department, University of
Wien, deposited in $mp\_ arc@math.utexas.edu$, \#94-66, to appear in
Journal of Statistical physics.

\item{[Ga2]} Gallavotti, G.: {\it Reversible Anosov diffeomorphisms and
large deviations.}, archived in $mp\_ arc@math.utexas.edu$, \#95-19.

\item{[Ga3]} Gallavotti, G.: {\it Topics in chaotic dynamics}, Lectures
at the Granada school, archived in $mp\_ arc@math.utexas.edu$, \#94-333.

\item{[Ga4]} Gallavotti, G.: {\it Coarse graining and chaos},
in preparation.

\item{[G]} Gibbs, J.W.: {\it Elementary principles in statistical
mechanics}, Ox Bow Press, 1981,(reprint).

\item{[HHP]} Holian, B.L., Hoover, W.G., Posch. H.A.: {\it Resolution of
Loschmidt's paradox: the origin of irreversible behavior in reversible
atomistic dynamics}, Physical Review Letters, {\bf 59}, 10--13, 1987.

\item{[LL]} Landau, L.D.,  Lifshitz, E.M.: {\it Fluid mechanics},
Pergamon press, Paris, 1959.

\item{[Lo]} Lorenz, E.: {\it Deterministic non periodic flow}, J.
of the Athmospheric Sciences, {\bf 20}, 130- 141, 1963.

\item{[LPR]} Livi, R., Politi, A., Ruffo, S.: {\it Distribution of
characteristic exponents in the thermodynamic limit}, Journal of
Physics, {\bf 19A}, 2033--2040, 1986.

\item{[Ma]} Marchioro, C.: {\it }, Communications in Mathematical
Physics, {\bf }, , .

\item{[P] } Pesin, Y.: {\it Dynamical systems with generalized hyperbolic
attractors: hyperbolic, ergodic and topological properties}, Ergodic
Theory and Dynamical Systems, {\bf 12}, p. 123--151, 1992.

\item{[PH]} Posch, H.A., Hoover, W.G.: {\it Non equilibrium molecular
dynamics of a classical fluid}, in "Molecular Liquids: new perspectives
in Physics and chemistry", ed. J. Teixeira-Dias, Kluwer Academic
Publishers, p. 527--547, 1992.

\item{[R1]} Ruelle, D.: {\it Chaotic motions and strange
attractors}, Lezioni Lincee, notes by S. Isola,
Accademia Nazionale dei Lincei, Cambridge University
Press, 1989; see also: Ruelle,
D.: {\it Measures describing a turbulent flow}, Annals of the New York
Academy of Sciences, {\bf 357}, 1--9, 1980.  For more technical
expositions see Ruelle, D.: {\it Ergodic theory of differentiable
dynamical systems}, Publications Math\'emathiques de l' IHES, {\bf 50},
275--306, 1980.

\item{[R2]} Ruelle, D.: {\it A measure associated with axiom A
attractors}, American Journal of Mathematics, {\bf98}, 619--654, 1976.

\item{[S1]} Sinai, Y.G.: {\it Dynamical systems with elastic reflections.
Ergodic properties of dispersing billiards}, Russian Mathematical
Surveys, {\bf 25}, 137--189, 1970.

\item{[S2]} Sinai, Y.G.: {\it Lectures in ergodic theory}, Lecture notes
in Mathematics, Prin\-ce\-ton U. Press, Princeton, 1977. See also
Sinai, Y.G.: {\it Markov partitions and $C$-diffeomorphisms}, Functional
Analysis and Applications, {\bf2}, 64--89, 1968, n.1; and
Sinai, Y.G.: {\it Construction of Markov partitions}, Functional analysis
and Applications, {\bf2}, 70--80, 1968, n.2.

\item{[SEM]} Sarman, S., Evans, D.J., Morriss, G.P.: {\it Conjugate pairing
rule and thermal transport coefficients}, Physical Review, {\bf 45A},
2233--2242, 1992.

\item{[UF]} Uhlenbeck, G.E., Ford, G.W.: {\it Lectures in Statistical
Mechanics}, American Mathematical society, Providence, R.I., pp. 5,16,30,
1963.

\ciao
\bye